\documentclass[preprint,11pt]{elsarticle}

\journal{Applied and Computational Harmonic Analysis}

\usepackage[latin1]{inputenc}
\usepackage[T1]{fontenc}
\usepackage{amssymb}
\usepackage{amsfonts}
\usepackage{amsmath}
\usepackage{mystyle}
\usepackage{url}

\usepackage{graphicx}
\usepackage{epsfig}
\usepackage{epstopdf}
\graphicspath{{./images/}}

\usepackage{vmargin}
\setpapersize{A4} 

\newcommand{\gab}[1]{}
\newcommand{\Jalal}[1]{}
\newcommand{\Charles}[1]{}
\newcommand{\Mline}[1]{}

\begin{document}

\begin{frontmatter}
\title{Sharp Support Recovery from Noisy\\Random Measurements by $\ell_1$ minimization}

\author[cd]{Charles Dossal} 
\ead{charles.dossal@math.u-bordeaux1.fr} 

\author[cd]{Marie-Line Chabanol}
\ead{Marie-Line.Chabanol@math.u-bordeaux1.fr} 

\author[gp]{Gabriel Peyr\'e} 
\ead{gabriel.peyre@ceremade.dauphine.fr} 

\author[jf]{Jalal Fadili} 
\ead{jalal.fadili@greyc.ensicaen.fr}

\address[cd]{IMB Universit\'e Bordeaux 1,\\351, cours de la Lib\'eration F-33405 Talence cedex, France}
\address[gp]{CNRS and CEREMADE, Universit\'e Paris-Dauphine,\\Place du Mar\'echal De Lattre De Tassigny, 75775 Paris Cedex 16, France}
\address[jf]{GREYC, CNRS-ENSICAEN-Universit\'e Caen,\\6 Bd du Mar\'echal Juin 14050 Caen Cedex, France}

\tnotetext[t1]{This work is supported by ANR grant NatImages ANR-08-EMER-009.}

\begin{abstract}
In this paper, we investigate the theoretical guarantees of penalized $\lun$-minimization (also called Basis Pursuit Denoising or Lasso) in terms of sparsity pattern recovery (support and sign consistency) from noisy measurements with non-necessarily random noise, when the sensing operator belongs to the Gaussian ensemble (i.e. random design matrix with i.i.d. Gaussian entries). More precisely, we derive sharp non-asymptotic bounds on the sparsity level and (minimal) signal-to-noise ratio that ensure support identification for most signals and most Gaussian sensing matrices by solving the Lasso with an appropriately chosen regularization parameter.  

Our first purpose is to establish conditions allowing exact sparsity pattern recovery when the signal is strictly sparse. Then, these conditions are extended to cover the compressible or nearly sparse case. In these two results, the role of the minimal signal-to-noise ratio is crucial. Our third main result gets rid of this assumption in the strictly sparse case, but this time, the Lasso allows only partial recovery of the support. We also provide in this case a sharp $\ldeux$-consistency result on the coefficient vector.

The results of the present work have several distinctive features compared to previous ones. One of them is that the leading constants involved in all the bounds are sharp and explicit. This is illustrated by some numerical experiments where it is indeed shown that the sharp sparsity level threshold identified by our theoretical results below which sparsistency of the Lasso solution is guaranteed meets the one empirically observed.
\end{abstract}


\end{frontmatter}


\Jalal{
\begin{itemize}
\item Il y a parfois une ambiguite sur la notion de non-asymptotique dans les resultats, et qui est donc a tirer au clair. En effet, on affirme que nos resultats son non-aysmptotiques (ce que j'ai renforce dans les contributions), mais ce n'est pas tjrs correctement justifie. Autant dans la cas du recovery exact (sparse et compressible), la probabilite est explicite, meme si elle n'est pas optimale. En revanche, dans le cas du recovery partiel sans condition sur le SNR minimal, dans la preuve, il y a bcp de w.o.p. sans expliciter justement la probabilite $P_2$. Je pense qu'il faut clarifier les choses. 

\item Encore sur ces probabilites, les expressions donnees dans le cas du recovery exact sont a verifier et clarifier. En effet, en non-aysmptotique, elle ne sont pas valables pour tout l'intervalle $[0,1)$ sur $a$ et $b$. Voir commentaire detaille dans le texte. Par ailleurs, ces probas semblent moins bonnes que celles de Candes et Plan.

\end{itemize}
}

\section{Introduction}
\label{sec:intro}

\subsection{Problem setup}
\label{subsec:setup}
The conventional wisdom in digital signal processing is the Shannon sampling theorem valid for bandlimited signals. However, such a sampling scheme excludes many signals of interest that are not necessarily bandlimited but can still be explained either exactly or accurately by a small number of degrees of freedom. Such signals are termed sparse signals. 

In fact we distinguish two types of sparsity: strict and weak sparsity (the latter is also termed compressibility). A signal $x$, considered as a vector in a finite dimensional subspace of $\mathbb{R}^p$, is strictly or exactly sparse if all but a few of its entries vanish; i.e.,\ if its support $I(x)=\supp{x}=\{1 \leq i \leq p \ \ | \ \ x[i] \neq 0\}$ is of cardinality $k \ll p$. A $k$-sparse signal is a signal where exactly $k$ samples have a non-zero value. Signals and images of practical interest may be {\em compressible} or {\em weakly sparse} in the sense that the sorted magnitudes $\abs{x^{\mathrm{sorted}}{[i]}}$ decay quickly. Thus $x$ can be well-approximated as $k$-sparse up to an error term (this property will be used when we will tackle compressible signals). If a signal is not sparse in its original domain, it may be {\em sparsified} in an appropriate 
orthobasis $\Phi$ (hence the importance of the point of view of computational harmonic analysis and approximation theory). Without loss of generality, we assume throughout that $\Phi$ is the standard basis.

The compressed sensing/sampling \cite{candes-robust,candes-near-optimal,donoho-cs} asserts that sparse or compressible signals can be reconstructed with  theoretical guarantees from far fewer measurements than the ambient dimension of the signal. Furthermore, the reconstruction is stable if the measurements are corrupted by an additive bounded noise. The encoding (or sampling) step is very fast since it gathers $n$ non-adaptive linear measurements that preserve the structure of the signal $x_0$: 
\eql{
\label{eq:obs}
	y = A x_0 + w \in \mathbb{R}^n,
}
where $A \in \mathbb{R}^{n \times p}$ is a rectangular measurement
matrix, i.e., $n < p$, and $w$ accounts for possible noise with
bounded $\ldeux$ norm. In this work, we do not need $w$ to be random and we consider that $A$ is drawn from the Gaussian matrix ensemble\footnote{In a statistical linear regression setting, we would speak of a random Gaussian design.}, i.e.,\ the entries of $A$ are independent and identically distributed (i.i.d.) $\mathcal{N}(0,1/n)$. The columns of $A$ are denoted $a_i$, for $i=1,\cdots,p$. In the sequel, the sub-matrix $\bA$ is the restriction of $A$ to the columns indexed by $I(x)$. To lighten the notation, the dependence of $I$ on $x$ is dropped and should be understood from the context.

The signal is reconstructed from this underdetermined system of linear equations by solving a convex program of the form:
\eql{
\label{eq:l1decoder}
x \in \uargmin{x \in \mathbb{R}^p} ~ \normu{x} \st A x - y \in \cC ~,
}
where $\cC$ is an appropriate closed convex set, and $\norm{x}_q:=\left(\sum_i \abs{x[i]}^q\right)^{1/q}$, $q \geq 1$ is the $\lp$-norm of a vector with the usual adaptation for $q=\infty$: $\normi{x}=\max_i \abs{x[i]}$. We also denote $\normz{x}$ as the $\lzero$ pseudo-norm which counts the number of non-zero entries of $x$. Obviously, $\normz{x}=\abs{I(x)}$. For any vector $x$, the notation $\overline{x} \in \RR^{\abs{I(x)}}$ means the restriction of $x$ to its support.

Typically, if $\cC=\{0\}$ (no noise), we end up with the so-called Basis Pursuit \cite{chen-basis-pursuit} problem
\begin{equation}
	\label{eq:BPl1}\tag{BP}
	\umin{x \in \RR^p} \normu{x} \st y = A x ~.
\end{equation}
Taking $\cC$ as the $\ldeux$ ball of radius $\epsilon$, we have a noise-aware variant of BP
\eql{\label{eql1constraint}\tag{$\lun$-constrained}
	\umin{ x \in \RR^p } \normu{x} \st \normd{A x - y} \leq \epsilon
}
where the parameter $\epsilon > 0$ depends on the noise level $\normd{w}$. This constrained form can also be shown to be equivalent to the $\lun$-penalized optimization problem, which goes by the name of Basis Pursuit DeNoising \cite{chen-basis-pursuit} or Lasso in the statistics community after \cite{TibshiraniLasso96}:
\eql{\label{eql1relax}\tag{Lasso}
	\umin{x \in \RR^p} \frac{1}{2}\normd{y - A x}^2 + \ga \normu{x} ~,
}
where $\ga$ is the regularization parameter. \eqref{eql1constraint} and \eqref{eql1relax} are equivalent in the sense that there is a bijection between $\ga$ and $\epsilon$ such that both problems share the same set of solutions. However, this bijection is unknown explicitly and depends on $y$ and $A$, so that in practice, one needs to use different algorithms to solve each problem, and theoretical results are stated using one formulation or the other. In this paper, we focus on the Lasso formulation.
It is worth noting that the Dantzig selector \cite{candes-dantzig,BickelLassoDantzig07} is also a special instance of \eqref{eq:l1decoder} when $\cC=\{z \in \mathbb{R}^p \big| \normi{\transp{A} z} \leq \gamma\}$.

The convex problems of the form \eqref{eql1constraint} and \eqref{eql1relax} are computationally tractable and many algorithms have been developed to solve them, and we only mention here a few representatives. Homotopy continuation algorithms \cite{osborne-homotopy,EfronLars,donoho-homotopy} track the whole regularization path. Many first-order algorithms originating from convex non-smooth optimization theory have been proposed to solve \eqref{eql1relax}. These include one-step iterative thresholding algorithms \cite{figueiredo-nowak-em,daubechies-iterated,bect-chambolle-iterative,combettes-proximal}, or accelerated variants \cite{figueiredo-grad-projection,bioucas-twist}, multi-step schemes such as \cite{nesterov-gradient} or \cite{beck-fista}. The Douglas-Rachford algorithm \cite{combettes-dr,Fadili09} is a first-order scheme that can be used to solve \eqref{eql1constraint}. A more comprehensive account can be found in \cite[Chapter 7]{StarckFadiliBook10}.

\subsection{Theoretical performance measures of the Lasso}
\label{subsec:perfmeasures}

These last years, we have witnessed a flurry of research activity where efforts have been made to investigate the theoretical guarantees of $\lun$ minimization by solving the Lasso for sparse recovery from noisy measurements in the underdetermined case $n < p$. Overall, the derived conditions hinge on strong assumptions on the structure and interaction between the variables in $A$ as indexed by $x_0$. An overview of the literature pertaining to our work will be covered in Section~\ref{subsec:overview} after notions are introduced so that the discussions are clearer.


Let $x_0$ be the original vector as defined in \eqref{eq:obs}, $f_0=A x_0$ the noiseless measurements, $x(\gamma)$ a minimizer of the Lasso problem and $f(\gamma)=Ax(\gamma)$.

\paragraph{Consistency} $\lp$-consistency on the signal $x$ means that the $\lp$-error $\norm{x_0-x(\gamma)}_q$, for typically $q=1$, $2$ or $\infty$, between the unknown vector $x_0$ and a solution $x(\gamma)$ of either \eqref{eql1relax} or \eqref{eql1constraint} comes within a factor of the noise level. 

\gab{J'ai enleve le oracle-type inequalities}

\paragraph{Sparsistency} Sparsity pattern recovery (also dubbed sparsistency for short or variable selection in the statistical language) requires that the indices and signs of the solutions $x(\gamma)$ are equal to those of $x_0$ for a well chosen value of $\ga$. Partial support recovery occurs when the recovered support is included (strictly) in that of $x_0$ with the correct sign pattern.
\\

In general, it is not clear which of these performance measures is better to characterize the Lasso solution. Nevertheless, in the noisy case, consistency does not tell the whole story and there are many applications where bounds on the $\lp$-error are insufficient to characterize the accuracy of the Lasso estimate. In this case, exact or partial recovery of the support, hence of the correct model variables, is the desirable property to have. Among other advantages, this allows for instance to circumvent the bias of the Lasso and thus enhance the estimation of $x_0$ and $A x_0$ using a debiasing procedure: recover the support $I$ by solving the Lasso, followed by least-squares regression on the selected variables $(a_i)_{i \in I}$; see e.g. \cite{candes-dantzig,CandesPlan09}. Our work falls within this scope and focuses on exact and partial support identification for both strictly sparse and compressible signals in the presence of noise on Gaussian random measurements.

\subsection{Literature overview}
\label{subsec:overview}

The properties of the Lasso have been extensively studied, including consistency and distribution of its estimates. There is of course a huge literature on the subject, and covering it fairly is beyond the scope of this paper. In this section, we restrict our overview to those works pertaining to ours, i.e., sparsity pattern recovery in presence of noise.

Much recent work aims at understanding the Lasso estimates from the point of view of sparsistency. This body of work includes \cite{CandesPlan09,candes-dantzig,donoho-stable-recovery,Meinshausen06,Greenshtein06,tropp-just-relax,wainwright-sharp-thresh,ZhaoYu06,Zou06}. For the Lasso estimates to be close to the model selection estimates when the data dimensions $(n,p)$ grow, all the aforementioned papers assumed a sparse model and used various conditions that require the irrelevant variables to be not too correlated with the relevant ones.

\paragraph{Mutual coherence-based conditions} Several researchers have studied independently the qualitative performance of the Lasso for either exact or partial sparsity pattern recovery of sufficiently sparse signals under a mutual coherence condition on the measurement matrix $A$; see for instance \cite{donoho-stable-recovery,fuchs-bounded-noise,tropp-just-relax,Bunea08} when $A$ is deterministic, and \cite{zhou-privacy} when $A$ is Gaussian.
However, mutual coherence is known to lead to overly pessimistic sparsity bounds.

\paragraph{Support structure-based conditions} These sufficient recovery conditions were refined by considering not only the cardinality of the support but also its structure, including the signs of the non-zero elements of $x_0$. Such criteria use the interactions between the relevant columns of $\bA = (a_i)_{i \in I}$ and the irrelevant ones $(a_i)_{i \notin I}$. More precisely, we define the following condition developed in \cite{fuchs-redundant-bases} to analyze the properties of the Lasso. This condition goes by the name of irrepresentable condition in the statistical literature; see e.g. \cite{ZhaoYu06,CandesPlan09,wainwright-sharp-thresh,Meinshausen09} and \cite{vandeGeer09} for a detailed review.

\begin{definition}
Let $I$ be the support of $x_0$ and $I^c$ its complement in $\{1,\cdots,p\}$. The irrepresentable (or Fuchs) condition is fulfilled if
\begin{align} 
      \label{eq-fuchs}
      F(x_0) &:= \normi{\transp{A_{I^c}}\bA (\bAt \bA)^{-1}\sign{\overline{x_0}}} = \max_{i \in I^c} |\ps{a_i}{d(x_0)}| < 1, \\
      \label{eq-dx}
      \qwhereq d(x_0) &:= \bA (\bAt \bA)^{-1} \sign{\overline{x_0}} ~.
\end{align}
\end{definition}
\Mline{peut-etre une question stupide, mais c'est une definition de la condition d'irrepresentabilite
  (auquel cas il faudrait le mettre dans la def) ou de F (auquel cas
  la condition ne fait pas partie de la def) ?} 

Condition \eqref{eq-fuchs} will also be the soul of our analysis in this paper. \\

The criterion \eqref{eq-fuchs} is closely related to the exact recovery coefficient (ERC) of Tropp \cite{tropp-just-relax}:
\begin{align} 
      \label{eq-erc}
      \mathrm{ERC}(x_0) := 1 - \max_{i \in I^c} \normu{(\bAt  \bA)^{-1}\bAt a_i} ~.
\end{align}
In \cite[Corollary 13]{tropp-just-relax}, it is established that if $\mathrm{ERC}(x_0) > 0$, then the support of the Lasso solution with a large enough parameter $\gamma$ is included in the one of the subset selection (i.e., $\lzero$-minimization) optimal solution.

In \cite{ZhaoYu06}, an asymptotic result is reported showing that \eqref{eq-fuchs}\footnote{In fact, a slightly stronger assumption requiring that all elements in \eqref{eq-fuchs} are uniformly bounded away from 1.} is sufficient for the Lasso to guarantee exact support recovery and sign consistency. It is also shown that \eqref{eq-fuchs} is essentially necessary for variable selection. \cite{Meinshausen06} develop very similar results and use similar requirements. \cite{Bach08} and \cite{nardi-lasso-asymp} derive asymptotic conditions for sparsistency of the block Lasso \cite{YuanLin06} by extending \eqref{eq-fuchs} and \eqref{eq-erc} to the group setting.


Reference \cite{CandesPlan09} proposes a non-asymptotic analysis with a sufficient condition ensuring exact support and sign pattern recovery of most sufficiently sparse vectors for matrices satisfying a weak coherence condition (of the order $(\log p)^{-1}$). Their proof relies upon \eqref{eq-fuchs} and a bound on norms of random sub-matrices developed in \cite{tropp-NormsRandom}. The work in \cite{wainwright-sharp-thresh} considers a condition of the form \eqref{eq-fuchs} to ensure sparsity pattern recovery. The analysis in that paper was conducted for both deterministic and standard Gaussian $A$ in a high-dimensional setting where $p$ and the sparsity level grow with the number of measurements $n$. That author also established that violation of \eqref{eq-fuchs} is sufficient for failure of the Lasso in recovering the support set. In \cite{omidiran-subset}, the sufficient bound on the number of measurements established in \cite{wainwright-sharp-thresh} for the standard Gaussian dense ensemble was shown to hold for sparse measurement ensembles. The works of \cite{CandesPlan09} and \cite{wainwright-sharp-thresh} are certainly the most closely related to ours. We will elaborate more on these connections by highlighting the similarities and differences in Section~\ref{subsec:relationpriorwork}. 


\paragraph{Variations on the Lasso} Other variations of the Lasso, such as the adaptive Lasso\footnote{The adaptive Lasso as seen in the statistical literature turns out to be a two-step procedure, where the second step is to solve a reweighted $\lun$ norm problem, with weights given by the Lasso estimate in the first step. In fact, this is a special case of the iteratively reweighted $\lun$-minimization \cite{candes-reweighted-l1}.} \cite{Zou06,Huang06} or multi-stage variable selection methods \cite{fan-overview-selection,Zhang09,Wasserman09,geer-thesh-adap-lasso,Meinshausen09}. For an overview of other penalized methods that have been proposed for the purpose of variable selection, see \cite{fan-overview-selection}.

\paragraph{Information-theoretic bounds} A recent line of research has developed information-theoretic sufficient and necessary bounds to characterize fundamental limits on \textit{minimal} signal-to-noise ratio (SNR), the number of measurements $n$, and tolerable sparsity level $k$ required for exact or partial support pattern recovery of exactly sparse signals by any algorithm including the optimal exhaustive $\lzero$ decoder \cite{wainwright-info-limits,fletcher-sparse-pattern,akcakaya-shannon,Reeves08,Wang2010,Aeron2010,Saligrama2010,Reeves10,hormati-estimation,Tune09,rad-sharp-pattern}. In most of these works, the bounds are asymptotic, i.e., they provide asymptotic scaling and typically require that the sparsity level $k$ varies at some rate (linearly or sub-linearly) with the signal dimension $p$ when $n$ grows to infinity. It is worth mentioning that a careful normalization is needed, for instance of the sampling matrix and noise, when comparing these results in the literature.

The paper \cite{wainwright-info-limits} was the first to consider the information-theoretic limits of exact sparsity recovery from the Gaussian measurement ensemble, explicitly identifying the minimal SNR (or equivalently $T = \min_{ i \in I(x_0) } |x_0[i]|$) as a key parameter. This analysis yielded necessary and sufficient conditions on the tuples $(n,p,k,T)$ for asymptotically reliable sparsity recovery. This complements the analysis of \cite{wainwright-sharp-thresh} by showing that in the sub-linear sparsity regime, i.e. $k = o(p)$, the number of measurements required by the Lasso\footnote{The shorthand notation $f \gtrsim g$ means that $g=O(f)$.} $n \gtrsim k \log (p - k)$ achieves the information-theoretic necessary bound. 

Subsequent work of \cite{fletcher-sparse-pattern,akcakaya-shannon,Reeves08,Wang2010,Aeron2010,Saligrama2010,Reeves10,hormati-estimation,Tune09,rad-sharp-pattern} has extended or strengthened this type of analysis to other settings (e.g. partial support recovery, other matrix ensembles, other scaling regimes, compressible case). 

\subsection{Contributions}
Most of the results developed in the literature on sparsistency of the
Lasso estimate exhibit asymptotic scaling results in terms of the
triple $(n,p,k)$, but this does not tell the whole story. One often
needs to know explicitly the exact numerical constants involved in the
bounds, not only their dependence on key quantities such as the SNR
and/or other parameters of the signal $x_0$.
As a consequence, the majority of sufficient conditions are more conservative than those suggested by empirical evidence.

In this paper, we investigate the theoretical properties of the Lasso estimate in terms of sparsity pattern recovery (support and sign consistency) from noisy measurements --the noise being not necessarily random-- when the measurement matrix belongs to the Gaussian ensemble. We provide precise {\textit{non-asymptotic}} bounds, including explicit sharp leading numerical constants, on the key quantities that come into play (sparsity level for a given measurement budget, minimal SNR, regularization parameter) to ensure exact or partial sparsity pattern recovery for both strictly sparse and compressible signals. Our results have several distinctive features compared to previous closely-connected works. This will be discussed in further details in Section~\ref{subsec:relationpriorwork}. Numerical evidence are reported in Section~\ref{sec-numerics} to confirm the theoretical findings.

\subsection{Organization of the paper}
The rest of the paper is organized as follows. We first state our main results and discuss the connections and novelties with respect to existing work. In Section~\ref{sec-exact-sparsity} and \ref{sec-compressibility}, we detail the proofs for exact recovery with strictly sparse and compressible signals, before proving the partial support recovery result in Section~\ref{sec-PartialSupportRecovery}. Numerical experiments are carried out in Section~\ref{sec-numerics}. Section~\ref{sec-conclusion} includes a final discussion and some concluding remarks.

\section{Main results}
\label{sec-contributions}

Our first result Theorem~\ref{TheoBruit} establishes conditions allowing exact sparsity pattern recovery when the signal is strictly sparse. Then, these conditions are extended to cover the compressible case in Theorem~\ref{TheoNotSparse2}. In these two results, the role of the minimal SNR is crucial. Our third main result in Theorem~\ref{TheoPartialRecovery} gets rid of this assumption in the strictly sparse case, but this time, the Lasso allows only partial recovery of the support. We also provide in this case a sharp $\ldeux$-consistency result on the Lasso estimate.

The three theorems are stated following the same structure: suppose that $(x_0,w)$ fulfill some requirements formalized by a set $\mathcal{Y}$, then with overwhelming probability (\wop for short) on the choice of $A$, the Lasso estimate obeys some property $\mathcal{P}$. It should be noted that these theorems imply in particular that \wop on the choice of $A$, for {\textit{most}} vectors $(x_0,w)\in \mathcal{Y}$, the Lasso estimate satisfies property $\mathcal{P}$, whatever the probability measure used on the set $\mathcal{Y}$.

The proof of Theorem~\ref{TheoBruit} is given in Section~\ref{sec-exact-sparsity}. We prove its extension to compressible signals as stated in Theorem~\ref{TheoNotSparse2} in Section~\ref{sec-compressibility}. Both proofs capitalize on an implicit formula of the Lasso solution. The proof of Theorem~\ref{TheoPartialRecovery} given in Section~\ref{sec-PartialSupportRecovery} is quite different, since no such implicit formula is used directly.

\subsection{Exact Support Recovery with Strictly Sparse Signals}

\begin{theoreme}\label{TheoBruit}
Let $A\in\mathbb{R}^{n\times p}$ be a Gaussian matrix, i.e.\ its entries are i.i.d. $\mathcal{N}(0,1/n)$, $w \in \RR^n$ is such that $\normd{w} \leq \varepsilon$, $0 \leq \alpha,\beta < 1$ and $p>e^{\frac{1}{2(1-\sqrt{\beta})}}$. Suppose that $x_0 \in \RR^p$ obeys
\eql{\label{eq-sparsity-constr}
	\normz{x_0} = k \leq \frac{\alpha \beta n}{2\log p}
}
and
\eql{\label{eq-minsnr-constr}
	\umin{ i \in I } |x_0[i]|=T \geq \frac{\six\varepsilon}{\sqrt{1-\alpha}} \sqrt{\frac{2\log p}{n}} ~.
}
Solve the Lasso problem from the measurements $y = A x_0 + w$. Then with probability $P(n,p,\alpha,\beta)$ converging to 1 as $n$ goes to infinity, the Lasso solution $x(\gamma)$ with 
\eql{\label{eq-gamma-value}
	\gamma=\frac{\varepsilon}{\sqrt{1-\alpha}}\sqrt{\frac{2\log p}{n}}
} 
is unique and satisfies 
\begin{equation*}
	\supp{x(\gamma)}=\supp{x_0} 
	\qandq
	\sign{\overline{x(\gamma)}}=\sign{\overline{x_0}} ~.
\end{equation*}
\end{theoreme}

\vskip 12pt

The proof (see Section~\ref{sec-exact-sparsity}) provides an explicit
bound for $P(n,p,\alpha,\beta)$, showing in particular that $P(n,p,\alpha,\beta)$ is larger than 
\begin{equation*}
1-\frac{1}{2} e^{-0.7\sqrt{\log n}}-\frac{1}{2\sqrt{\pi\log p}} - o\left(\frac{1}{\log p}\right)-o(e^{-0.7\sqrt{\log n}}) ~,
\end{equation*}
although this bound on the probability is far from optimal.\\

In plain words, Theorem~\ref{TheoBruit} asserts that for $(\alpha,\beta)\in[0,1)$ 
the support and the sign of most vectors obeying \eqref{eq-sparsity-constr} can be recovered using the Lasso if the non-zero coefficients of $x_0$ are large enough compared to noise. This bound on the sparsity of $x_0$ turns out to be optimal, since for any $c>1$, for most vectors $x_0$
such that $\normz{x_0}\geq\frac{c n}{2\log p}$, the support cannot be recovered using the Lasso even with no noise. Indeed, \cite{fuchs-redundant-bases} and \cite{dossal-topological} proved that the Lasso solution for any $\gamma$ shares the same sign and the same support as $x_0$ when $y=Ax_0$ if and only if
\begin{equation*}
	\max_{j\notin I}|
		\ps{a_j}{\bA(\bAt \bA)^{-1}\sign{\overline{x_0}}}|\leq 1 ~.
\end{equation*}  
Note in passing the difference with the strict inequality in \eqref{eq-fuchs}.
On the other hand, if $\normz{x_0}\geq\frac{c n}{2\log p}$ with $c > 1$, then \wop 
$\normd{\bA(\bAt \bA)^{-1}\sign{\overline{x_0}}}^2\geq
\frac{C n}{2\log p}$ for some $C > 1$ and sufficiently large $p$. As a result, $\max_{j\notin
  I}|\ps{a_j}{\bA(\bAt \bA)^{-1}\sign{\overline{x_0}}}|\geq \sqrt{C}>1$.
This informal optimality discussion is consistent with the information-theoretic bounds of \cite{wainwright-info-limits}, where it was proved that the number of measurements required by the Lasso achieves the (asymptotic) information-theoretic necessary bound that has the scaling \eqref{eq-sparsity-constr} when the sparsity regime is sub-linear and $T^2 \sim 1/\normz{x_0}$.

An important feature of Theorem~\ref{TheoBruit} is that all the
constants are made explicit and are governed by the two numerical
constants $\alpha$ and $\beta$. The role of $\alpha$ is very
instructive since when lowering $\gamma$ by decreasing $\alpha$, the threshold on the minimal SNR is decreased to allow smaller coefficients to be recovered, but simultaneously the probability of success gets lower and the number of measurements required to recover the $k$-sparse signal increases. The converse applies when $\alpha$ is increased. On the other hand, increasing $\beta$ (in an appropriate range; see Section~\ref{subsec-cond-c2} for details) allows a higher threshold on the sparsity level, but again at the price of a smaller probability of success.

\newcommand{\xk}{x^k}

\subsection{Support Recovery with Compressible Signals}

\gab{J'ai reformulé le théorème avec l'approximation $k$ terme. J'ai aussi enlevé la remarque avec les weak Lp, en disant juste que $x_0-\xk$ est petit. Parler de weak Lp en dimension finie me semble difficile. }

Theorem~\ref{TheoBruit} can be easily extended to weakly sparse or compressible signals. We consider the best $k$-term approximation $\xk$ of $x_0$ obtained by keeping only the $k$ largest entries from $x_0$ and setting the others to zero. Obviously, $k = |I(\xk)|$. This is equivalently defined using a thresholding
\eql{\label{eq-dfn-xk}
	\xk[i] = \choice{
		x_0[i] \quad\text{if}\quad |x_0[i]| \geq T,\\
		0 \quad \text{otherwise.}
	}
}
A signal is generally considered as compressible if the residual $\xk-x_0$ is small. For sparsistency to make sense in this compressible case, additional assumptions are required, namely that the largest components $\xk$ of the signal are significantly larger than the residual $\xk-x_0$. This is made formal in the following theorem.


\begin{theoreme}\label{TheoNotSparse2}
	Let $A$, $\alpha$, $\beta$ and $p$ as in Theorem~\ref{TheoBruit}. 
	We measure $y=A x_0 + w$, and let $\xk$ be the best $k$-term
        approximation of $x_0$ where $k$ satisfies
        \eqref{eq-sparsity-constr}. We denote 
\eq{
\Delta =  \frac{2}{\sqrt{1+2\sqrt{\alpha}-3\alpha}}\sqrt{\frac{2\log p}{n}}.
}
Suppose that  
	\eql{\label{eq-th-compress-epsilon}
		\normd{w} + 4\normd{x_0-\xk} \leq \varepsilon,
	}
	$T$ as defined in \eqref{eq-dfn-xk} is such that
	\eql{\label{eq-th-compress-1}
          T \geq 5.5 \Delta \varepsilon
	}
	and
	\eql{\label{eq-th-compress-2}
          \normi{x_0-\xk} \leq \frac{4}{5}(1-\sqrt
		\alpha) \Delta \varepsilon.
	}
	Then, with probability $P_2(n,p,\alpha,\beta)$ converging to 1 as $n$ goes to infinity, 
	the solution $x(\gamma)$ of the Lasso from measurements $y$ with 
	\eql{\label{eq-th-compress-gamma}	
          \gamma = \Delta \varepsilon
	}
	is unique and satisfies 
	\eq{
		\supp{x(\gamma)}=\supp{\xk} \qandq 
		\sign{\overline{x(\gamma)}}=\sign{\overline{\xk}} ~.
	}
\end{theoreme}

\vskip 12pt

Again, all the leading constants are explicit. Conditions \eqref{eq-th-compress-1} and \eqref{eq-th-compress-2} impose compressibility constraints on the signal, namely that the magnitude of the $k$ largest components of $x_0$ are well above the average magnitude $\varepsilon/\sqrt{n}$ of the residual, and that the latter is ``flat'', since the ratio of its $\linf$ and $\ldeux$ norms should be small. 

The proof (see Section~\ref{sec-compressibility}) provides an explicit
bound for $P_2(n,p,\alpha,\beta)$, showing that $P_2(n,p,\alpha,\beta)$ is greater than 
\begin{equation*}
1-\frac{1}{2} e^{-0.7\sqrt{\log n}}-\frac{1}{2\sqrt{\pi\log p}} - o\left(\frac{1}{\log p}\right)-o(e^{-0.7\sqrt{\log n}}) ~,
\end{equation*}
although once again this bound on the probability is far from optimal.\\

Theorem~\ref{TheoNotSparse2} encompasses the strictly sparse case, Theorem \ref{TheoBruit}, which is easily recovered by letting $x_0=\xk$. The parameter $\alpha$ plays a similar role in both theorems. Furthermore, in Theorem \ref{TheoNotSparse2}, the Lasso solution becomes more tolerant to compressibility errors $x_0-\xk$ as $\alpha$ decreases. This however comes at the price of a lower probability of success as indicated in our proof.

\subsection{Partial Support Recovery with Strictly Sparse Signals}
In both previous theorems, the assumption on $T$ plays a pivotal role: if $T$ is too small, there is no way to distinguish the small components of $x_0$ from the noise; see also the discussion and literature review in Section~\ref{subsec:overview}. Nevertheless, if no assumptions are made on $T$, one can nevertheless expect to partly recover the support of $x_0$. This is formalized in the following result.

\begin{theoreme}\label{TheoPartialRecovery}
Let $A$, $\alpha$ and $\beta$ as in Theorem~\ref{TheoBruit}. We measure $y=Ax_0+w$, where $x_0$ fulfills \eqref{eq-sparsity-constr}.
Then with probability $P_3(n,p,\alpha,\beta)$ converging to 1 as $n$ goes to infinity, the solution $x(\gamma)$ of the Lasso form measurements $y$ with 
\[
\gamma=\frac{\varepsilon}{\sqrt{1-\alpha}}\sqrt{\frac{2\log p}{n}}
\]  
is unique and satisfies 
\[
\supp{x(\gamma)}\subset \supp{x_0}.
\] 
Moreover, the Lasso solution is $\ldeux$-consistent: 
\eql{\label{eq-th-partial-l2}
	\normd{x_0-x(\gamma)}\leq \left(2+\sqrt{\frac{\al}{1-\al}}\right) \: \varepsilon ~.
}
\end{theoreme} 

The proof in Section~\ref{sec-PartialSupportRecovery} provides an explicit lower bound for $P_3(n,p,\alpha,\beta)$, and shows that $P_3(n,p,\alpha,\beta)$ is larger than 
\begin{equation*}
1-e^{-\frac{n\left(1-\sqrt{\beta}-\sqrt{\frac{k}{n}}\right)^2}{2}}-\frac{1}{2\sqrt{\pi\log p}} ~.
\end{equation*}
As before, this bound on the probability is not optimal.\\
\vskip 12pt

If $\gamma$ is large enough it is clear that $\supp{x(\gamma)}\subset \supp{x_0}$ since for $\gamma\geq \normi{\transp{A}y}$, $x(\gamma)=0$. Theorem~\ref{TheoPartialRecovery} provides a parameter $\gamma$ proportional to $\varepsilon$ that ensures a partial support recovery without any assumption on $T$. It also gives a sharp upper bound on $\ldeux$-error of the Lasso solution. This result remains valid under the additional hypotheses of Theorem~\ref{TheoBruit} or \ref{TheoNotSparse2} allowing exact recovery of the support.

\subsection{Connections to related works}
\label{subsec:relationpriorwork}

\paragraph{Sparsistency}
As we mentioned in Section~\ref{subsec:overview}, our work is closely related to \cite{CandesPlan09,wainwright-sharp-thresh}, but is different in many important ways that we summarize as follows.
 
\begin{itemize}
\item Deterministic vs random measurement matrices: the work of \cite{CandesPlan09} considers deterministic matrices satisfying a weak incoherence condition. Our work focuses on the classical Gaussian ensemble.
\gab{J'ai enleve la phrase sur le fait que Wainright considere des matrices deterministes}

\item Asymptotic vs non-asymptotic analysis: the analysis in \cite{wainwright-sharp-thresh} applies to high-dimensional setting where even the sparsity level $k$ grows with the number of measurements $n$. As a result, $k$ appears in the statements of the probabilities, which thus requires that $k \to +\infty$. This is very different from our setting as well as that of \cite{CandesPlan09} where the probabilities depend solely on the dimensions of $A$. We believe that this is more natural in many applications.

\item Random vs deterministic noise: in both previous works, the noise is stochastic (Gaussian in \cite{CandesPlan09} and sub-Gaussian in \cite{wainwright-sharp-thresh}). In our work, we handle any noise with a finite $\ldeux$-norm.


\item Leading numerical constants: these are not always explicit and sharp in those works. 
The constant involved in the sparsity level upper-bound in \cite[Theorem 1.3]{CandesPlan09} is not given, whereas \eqref{eq-sparsity-constr} gives an explicit and sharp bound.
The bounds \eqref{eq-minsnr-constr} and \eqref{eq-gamma-value} on $T$ and $\gamma$ are similar to those given in \cite[Theorem 1.3]{CandesPlan09} once specialized for $\al=3/4$.
In \cite[Theorem 2]{wainwright-sharp-thresh}, the constant appearing in the lower-bound on $T$ is not given, whereas \eqref{eq-minsnr-constr} provides an explicit expression that is shown to be reasonably good in Section \ref{sec-numerics}.
\Charles{Je serai favorable à virer les lignes suivantes et à les garder pour les reviewers s'ils les demandent.} 

\item Compressible signals: to the best of our knowledge, the compressible case has not been covered in the literature, and Theorem~\ref{TheoNotSparse2} appears then as a distinctively novel result of this paper. 


\item $\ldeux$-consistency: such a result is not given in those references. A bound on the $\ldeux$-prediction error on $A x_0 - A x(\gamma)$ is proved in \cite{CandesPlan09}. An $\linf$-consistency is established in \cite{wainwright-sharp-thresh}, which is an immediate consequence of sparsistency. Our method of proof differs significantly from the one used in \cite{wainwright-sharp-thresh}, and in particular it naturally leads to the  $\ldeux$-consistency result. 

\item Exact and partial support recovery: in \cite{CandesPlan09} the partial recovery case was not considered. \Charles{Est ce que la precision suivante est vraiment utile ?} In \cite{wainwright-sharp-thresh}, exact and partial recovery are somewhat handled simultaneously, while we give two distinct results for each case. 

\end{itemize}

\paragraph{$\ldeux$-consistency}
This property of the Lasso estimate has been widely studied by many authors under various sufficient conditions. Theorem~\ref{TheoPartialRecovery} may then be compared to this literature, and we here focus on results based on the restricted isometry property (RIP) \cite{candes-decoding} and more or less similar variants in the literature; see the discussion in \cite{Meinshausen09} and the review in \cite{vandeGeer09}. 

The RIP results are uniform and ensure $\ldeux$-stability of the Lasso estimate for {\textit{all}} sufficiently sparse vectors from noisy measurements, whereas Theorem \ref{TheoPartialRecovery} guarantees that the Lasso estimate is $\ldeux$-consistent for {\textit{most}} sparse vectors and a given matrix. When $A$ is Gaussian, the scaling of the sparsity bound is $O(n/\log(p/n))$ for RIP-based results which is better than $O(n/\log p)$ in Theorem~\ref{TheoPartialRecovery}. Note that the scaling $O(n)$ was derived in \cite{donoho-for-most-approx} when $A$ belongs to the uniform spherical ensemble to ensure $\ldeux$-stability of the Lasso estimate for most matrices $A$, although the leading constants are not given explicitly. However, the RIP is a worst-case analysis, and the price is that the leading constants in the sufficient sparsity bounds are overly small. In contrast, the leading numerical constants in our sparsity and $\ldeux$-consistency upper-bounds are explicit and solely controlled by $(\alpha,\beta) \in [0,1)^2$. For instance, it can be verified from our proof that the value of the sparsity upper-bound we provide is actually larger than the bounds obtained from the RIP for $p$ up to $e^{100}$. Finally, the RIP is a deterministic property that turns out to be satisfied by many ensembles of random matrices other than the Gaussian. Our Theorem~\ref{TheoPartialRecovery} could presumably be extended to sub-Gaussian matrices (e.g. using \cite[Corollary V.2.1]{FeldheimSodin10}), but this needs further investigation that we leave for a future work.

\section{Proof of Support Identification of Exactly Sparse Signals}
\label{sec-exact-sparsity}

This section gives the proof of Theorem~\ref{TheoBruit}. Recall that $\bar{x}$ is the restriction of $x$ to its support $I(x)$, and $A_I$ the corresponding sub-matrix. 
We also denote the Moore-Penrose pseudo-inverse of $\bA$ as
\eq{
	\bA^+ = (\bAt \bA)^{-1}\bAt.
}

\subsection{Optimality Conditions for Penalized Minimization}
\label{subsec_spars1}

From classical convex analysis, the first order optimality conditions show that a vector $x^\star$ is a solution of the Lasso if and only if
\begin{equation}\label{l1condF2}
	\choice{
		\bAt (y-Ax^\star)=\ga\sign{\ol{x^\star}} \\
		\forall j\notin I, \quad |\dotp{a_j}{y-Ax^\star}| \leq \ga,
	}
\end{equation}
where $I=I(x^\star)$.

Hence if the goal pursued is to ensure that $I(x^\star)=I(x_0)=I$ and $\sign{x^{\star}} =
\sign{x_0}$, the only candidate solution of the Lasso is
\begin{equation}\label{defxetoile}
	\ol{x^\star} = \ol{x_0}-\ga(\bAt \bA)^{-1}\sign{\ol{x_0}}+ \bA^+ w.
\end{equation} 
Consequently, a vector $x^\star$ is a solution of the Lasso if and only the two following conditions are met :
\begin{align}
	\sign{x_0} = \sign{x^\star} \tag{$C_1$} \\
	\forall j \notin I(x_0),\quad |\ps{a_j}{\ga d(x_0)+P_{\spanI^{\perp}}(w)}|\leq \ga \tag{$C_2$}
\end{align}
where $\spanI=\Span(\bA)$, $P_{\spanI^{\perp}}$ is the orthogonal projection on the subspace orthogonal to $\spanI$, and $d(x_0)$ is defined in \eqref{eq-dx}.

Sections \ref{subsec-cond-c1} and \ref{subsec-cond-c2} show that under the hypotheses of Theorem \ref{TheoBruit}, conditions $(C_1)$ and
$(C_2)$ are in force with probability converging to $1$ as $n$ goes to infinity. This will thus conclude the proof of Theorem \ref{TheoBruit}.

\subsection{Condition $(C_1)$}
\label{subsec-cond-c1}

To ensure that $\sign{x_0}=\sign{x^\star}$, it is sufficient that 
\eql{\label{eq-toprove-c1}
	\normi{\gamma(\bAt \bA)^{-1}\sign{\ol{x_0}} + \bA^+ w} \leq T ~.
}
We prove that this is indeed the case \wop. 

Lemma~\ref{LemmeBorneInf}, whose proof is given in Appendix~\ref{subsec-lemma-spectral-sup}, shows that 
$\ga=\frac{\varepsilon}{\sqrt{1-\alpha}}\sqrt{\frac{2\log p}{n}}\leq
\frac{T}{\six}$ implies 
\eq{
	\ga\normi{(\bAt \bA)^{-1}\sign{\ol{x_0}}} \leq \frac{T(1+4\sqrt{\alpha})}{\six}
}
with probability greater than $1-kp^{-1.28}-2e^{-\frac{n\alpha(0.75\sqrt{2}-1)^2}{4\log p}}$.


To prove \eqref{eq-toprove-c1}, we will now bound $\normi{\bA^+ w}$. To this end, we split it as follows
\eq{
	\normi{\bA^+ w} = D_1 \times D_2 \times D_3 \times \normd{w},
}
where 
\eq{
	D_1 = \frac{ \normi{\bA^+ w} }{\normd{\bA^+ w}}, \quad
	D_2 = \frac{\normd{\bA^+ w}}{\normd{\bAt w}}, \quad
	D_3 = \frac{\normd{\bAt w}}{\normd{w}}.
}

\paragraph{Bounding $D_1$} 
As $A$ and $w$ are independent, Lemma~\ref{lem-rotinv}, proved in Appendix~\ref{subsec-rotinv}, shows that the distribution of $\bA^+ w$ is invariant under orthogonal transforms on $\RR^k$. Therefore the random variable 
\eq{	
	\frac{\bA^+ w}{\normd{\bA^+ w}}
}
is uniformly distributed on the unit $\ldeux$ sphere of $\RR^k$.

Using the concentration Lemma~\ref{lem-unifsphere}, detailed in Appendix~\ref{subsec-lemma-concentration}, with $\epsilon= \left(\frac{8\log n \log k}{k^2} \right)^{\frac{1}{4}}$, it follows that  
\eqnl{\label{eq-bound-D1}
	P\left(D_1 \leq \sqrt{\frac{2}{k}}(2\log n\log k)^{\frac{1}{4}}\right)
	& \geq & 1-4ke^{-\sqrt{2\log n \log k}} \nonumber \\
        & \geq & 1-\max\left(4n^{-\frac{1}{3}},8e^{-\sqrt{2\log (2n) }}\right).
}

One can notice that  $D_1\leq 1$ actually gives a better bound if $k$ is small compared to $n$. Moreover the bound on the probability is $1-4n^{-\frac{1}{3}}$ for $k$ big.

\paragraph{Bounding $D_2$} 


$D_2$ is bounded by the maximum of the eigenvalue of $(\bAt \bA)^{-1}$. Indeed, owing to Lemma~\ref{LemmeVSWishart} with $t=1-\sqrt\frac{k}{n} - {2^{-\frac{1}{8}}}$, we arrive at
\eql{\label{eq-bound-D2}
        P\left(D_2 \leq 2^{\frac{1}{4}}\right)\geq 1-e^{-\frac{n}{2}\left(1-{2^{-\frac{1}{8}}}-\frac{1}{\sqrt{2\log p}}\right)^2} ~.
}

\paragraph{Bounding $D_3$} 

Let's write
\eq{
	D_3^2 = \frac{1}{\normd{w}^2} \sum_{i \in I} | \dotp{a_i}{w} |^2.
}
Since each  $\ps{a_i}{w}$ is a zero-mean Gaussian variable with variance $\frac{\normd{w}^2}{n}$, 
the variable
\eq{
	\frac{n\normd{\bAt w}^2}{\normd{w}^2},
}
follows a $\chi^2$ distribution with $k$ degrees of freedom. Therefore, in virtue of the concentration Lemma~\ref{LemmeBorneChi2}, stated in Appendix~\ref{subsec-lemma-concentration}, applied with 
\eq{
	1+\delta=2\sqrt{\frac{\log n}{\log k}}
}
we obtain
\eq{
	P\left(D_3^2 \leq \frac{2k\sqrt{\log n}}{n\sqrt{\log k}}\right)
	\geq 1-\frac{1}{\sqrt{2\pi k}} e^{-k\left(\sqrt{\frac{\log n}{\log k}}-\frac{1}{2} - \frac{\log 2}{2} - \frac{1}{4} \log\left(\frac{\log n}{\log k}\right)\right)} 	\geq 1-\frac{1}{2}e^{-0.7\sqrt{\log n}}
}
This last bound may be pessimistic; when $k$ is large this probability is actually much bigger.
This shows that \wop,
\eql{\label{eq-bound-D3}
	D_3 \leq \sqrt{\frac{2k}{n}}\left(\frac{\log n}{\log k}\right)^\frac{1}{4}.
}


Putting \eqref{eq-bound-D1}, \eqref{eq-bound-D2} and \eqref{eq-bound-D3}, we conclude that 
\eql{\label{eq-bound-Apw}
	\normi{\bA^+ w} \leq 2\varepsilon\sqrt{\frac{2\log n}{n}},
}
with probability greater than 
\eq{
1-\frac{1}{2} e^{-0.7\sqrt{\log
    n}}-e^{-\frac{n}{2}\left(1-2^{-\frac{1}{8}}-\frac{1}{\sqrt{2\log
        p}}\right)^2}- \max\left(4n^{-\frac{1}{3}},8e^{-\sqrt{2\log
      (2n) }}\right) -kp^{-1.28}-2e^{-\frac{n\alpha(0.75\sqrt{2}-1)^2}{4\log p}}}
which converges to $1$ as $n \to +\infty$.

In turn, the bound \eqref{eq-bound-Apw} becomes, under assumption \eqref{eq-minsnr-constr} on $T$, 
\eq{
	\normi{\bA^+ w} \leq \frac{2T\sqrt{1-\alpha}}{\six}.
}
This shows that condition $(C_1)$ is in force with probability converging to $1$ as $n \to +\infty$.

\subsection{Condition $(C_2)$}
\label{subsec-cond-c2}


Let's introduce the following vector
\eql{\label{eq-defn-u}
	u = \ga d(x_0) + P_{\spanI^{\perp}}(w),
}
which depends on both $x_0$ and $w$.

Clearly, to comply with $(C_2)$, we need to bound $(\ps{a_j}{u})_{j\notin I}$ \wop. We will start by bounding $\normd{u}$.

\paragraph{Bounding $\normd{u}$}

As $d(x_0) \in \spanI$, the Pythagorean theorem yields
\eql{\label{eq-decom-pythagore}	
 	\normd{u}^2 = \ga^2 \normd{d(x_0)}^2 + \normd{P_{\spanI^{\perp}}(w)}^2.
}

Let $S = \sign{\ol{x_0}}$. Then 
\eq{
	\frac{nk}{\normd{d(x_0)}^2} = \frac{n\normd{S}^2}{ \transp{S} (\bAt \bA)^{-1}S}.
}
Since $x_0$ and $A$ are independent, Lemma~\ref{LemmeMuirhead}, stated in Appendix~\ref{subsec-lemma-concentration}, shows that $\frac{nk}{\normd{d(x_0)}^2}$ is $\chi^2$-distributed with $n-k+1$ degrees of freedom. Thanks to Lemma~\ref{LemmeBorneChi2bis}, see Appendix \ref{subsec-lemma-concentration}, it follows that for all $\delta>0$,
\eq{
	P\left(\frac{nk}{n-k+1}<(1-\delta)\normd{d(x_0)}^2\right)\leq e^{\frac{(n-k+1)\log(1-\delta)}{2}} ~.
}
Since $\frac{k}{n}\leq \frac{1}{2\log p}$, we obtain for $p \geq e^{\frac{1}{2\delta}}$, 
\eq{
	P\left(k<\normd{d(x_0)}^2(1-\delta)^2\right) \leq e^{\frac{n\log(1-\delta)(4-\delta)}{8}}.
}
Choosing $\de$ such that $(1-\delta)>\sqrt{\beta}$, we have
\eq{
	P\left(\normd{d(x_0)}^2\leq \frac{k}{\beta}\right) \geq 1-e^{\frac{n(3-\sqrt{\beta})\log \beta}{16}} ~.
}
This shows that
\eq{
	\normd{d(x_0)}^2 \leq \frac{k}{\beta}
}
with probability converging to $1$ as $n \to +\infty$.

It is worthy to mention that the condition $p>e^{\frac{1}{2(1-\sqrt{\beta})}}$ actually guarantees the existence of a suitable $\delta$.

As $P_{\spanI^{\perp}}$ is an orthogonal projector, we have $\normd{P_{\spanI^{\perp}}(w)} \leq
\normd{w}\leq \varepsilon$. Together with \eqref{eq-decom-pythagore}, this shows that 
\eql{\label{eq-bound-u}
	P\left(\normd{u}^2 \leq \ga^2\frac{k}{\beta}+\varepsilon^2\right)\geq 1-e^{\frac{n(3-\sqrt{\beta})\log \beta}{16}} ~.
}

\paragraph{Bounding $\max_{j \notin I} |\dotp{u}{a_j}|$}

For a fixed $u$, 
the random variables $\left(\ps{a_j}{u}\right)_{j \notin I}$ are zero-mean Gaussian variables with variance $\frac{\normd{u}^2}{n}$. 

Using the bound \eqref{eq-bound-u}, traditional arguments from the concentration of the maximum of Gaussian variables tell us that 
\eql{\label{eq-exact-C2-max}
	\max_{j\notin I}|\ps{a_j}{u}| \leq \sqrt{\frac{2\log p}{n} \left(\ga^2\frac{k}{\beta}+\varepsilon^2\right)}
}
with a probability larger than 
\[
1-e^{\frac{n(3-\sqrt{\beta})\log \beta}{16}}-\frac{1}{2\sqrt{\pi\log p}}.
\]
In turn, this implies that condition $(C_2)$ is in force \wop if 
\eq{
	\sqrt{\frac{2\log p}{n}\left(\ga^2\frac{k}{\beta}+\varepsilon^2\right)}\leq \ga.
}
This holds if
\eq{
	\frac{\varepsilon}{\sqrt{1-\alpha}}\sqrt{\frac{2\log p}{n}} \leq \ga.
}
This concludes the proof of Theorem \ref{TheoBruit}, and shows that
overall 
\begin{align*}
P(n,p,\alpha,\beta) & \geq 1-\frac{1}{2} e^{-0.7\sqrt{\log
    n}}-e^{-\frac{n}{2}\left(1-2^{-\frac{1}{8}}-\frac{1}{\sqrt{2\log
        p}}\right)^2}- \max\left(4n^{-\frac{1}{3}},8e^{-\sqrt{2\log
      (2n) }}\right)\\
& -kp^{-1.28}-2e^{-\frac{n\alpha(0.75\sqrt{2}-1)^2}{4\log p}} -
e^{\frac{n(3-\sqrt{\beta})\log \beta}{16}}-\frac{1}{2\sqrt{\pi\log
    p}}.
\end{align*}

\section{Proof of Support Identification of Compressible Signals}
\label{sec-compressibility}

To prove this theorem, we capitalize on the results of Section~\ref{subsec_spars1} by noting that $y=Ax^k+A(x_0-x^k)+w:=Ax^k+Ah + w$, and replacing  $x_0$ by $x^k$ and $w$ by $w_2 = Ah+w$. With these change of variables, it is then sufficient to check conditions $(C_1)$ and $(C_2)$ with the notable difference that the noise $w_2$ is not independent of $A$ anymore. More precisely, $w_2$ is independent of $(a_i)_{i\in I}$ but not of $(a_j)_{j\notin I}$. 

\paragraph{Condition $(C_1)$}
Since this condition only depends on $\bA$, it is verified with
probability converging to 1 as $n \to +\infty$, as in the proof of
Theorem~\ref{TheoBruit}, provided that $T\geq \six \gamma$ and $\normd{w_2}\leq \frac{T}{\six}
\sqrt{\frac{(1-\alpha)n}{2\log p}}$. The first condition is a direct
consequence of assumptions
\eqref{eq-th-compress-1} and \eqref{eq-th-compress-gamma}. 
Moreover,  $\normd{w_2} \leq \normd{w} + \normd{Ah}$, where $Ah$ is a zero-mean Gaussian vector, whose entries are independent with variance $\frac{\normd{h}^2}{n}$. Therefore $\frac{n \normd{Ah}^2}{\normd{h}^2}$ has a $\chi^2$ distribution with $n$ degrees of freedom. We then derive from the concentration Lemma~\ref{LemmeBorneChi2} that
\eq{
        P\left(\normd{Ah} \leq 2\normd{h}\right) \geq 1-\frac{1}{3\sqrt{2\pi n}} e^{-0.8n} ~.
}
Under assumptions \eqref{eq-th-compress-epsilon}-\eqref{eq-th-compress-1}, the last inequality implies that 
\eq{
\normd{w_2} \leq \normd{w} + 2\normd{h} \leq \varepsilon \leq
\frac{T}{\six \Delta} \leq \frac{T}{\six}
\sqrt{\frac{(1-\alpha)n}{2\log p}}
}
with probability that tends to 1 as $n \to +\infty$. Condition $(C_1)$
is thus satisfied with a probability larger than
\begin{align*}& 1-\frac{1}{2} e^{-0.7\sqrt{\log
    n}}-e^{-\frac{n}{2}\left(1-2^{-\frac{1}{8}}-\frac{1}{\sqrt{2\log
        p}}\right)^2}- \max\left(4n^{-\frac{1}{3}},8e^{-\sqrt{2\log
      (2n) }}\right)
      -kp^{-1.28}\\
&-2e^{-\frac{n\alpha(0.75\sqrt{2}-1)^2}{4\log p}}
-\frac{1}{3\sqrt{2\pi n}} e^{-0.8n}
.
\end{align*}

\paragraph{Condition $(C_2)$}
For any $j \notin I$, define the vector $v_j = w_2-h[j]a_j$. In particular, $v_j$ is independent of $a_j$. Condition $(C_2)$ now reads:
\eq{
	\forall j\notin I,  |\ps{a_j}{\gamma d(x^k)+P_{\spanI^{\perp}}(v_j)+h[j]P_{\spanI^{\perp}}(a_j)}| \leq \gamma ~,
}
where the vector $d(x^k)$ is defined replacing $x_0$ by $x^k$ in \eqref{eq-dx}.\\

\noindent
Similarly to \eqref{eq-bound-u}, it can be shown that \wop
\eq{
\normd{\ga d(x^k) + P_{\spanI^{\perp}}(v_j)}^2 \leq \ga^2\frac{k}{\beta}+\normd{v_j}^2 ~.
}
On the other hand, $\normd{v_j}\leq \normd{w_2}+\normi{h}\normd{a_j}$, and $n\normd{a_j}^2$ is $\chi^2$-distributed with $n$ degrees of
freedom. Applying Lemma~\ref{LemmeBorneChi2} to bound $\normd{a_j}$ by
$2$ for all $j$ and using similar arguments to those leading to \eqref{eq-exact-C2-max}, we get
\eq{
	\max_{j\notin I} |\ps{a_j}{\ga d(x^k) + P_{\spanI^{\perp}}(v_j)}| \leq 
	\sqrt{\frac{2\log p}{n}
        \left(\ga^2\frac{k}{\beta}+(\normd{w}+4\normd{h})^2\right)}
}
with probability larger than $1-\frac{p+1}{3\sqrt{2\pi n}} e^{-0.8n}-\frac{1}{2\sqrt{\pi\log p}}$,
converging to 1 as $n \to +\infty$. It then follows from assumptions \eqref{eq-th-compress-epsilon} and \eqref{eq-th-compress-gamma} that \wop
\eql{\label{compres1}
	\max_{j\notin I} |\ps{a_j}{\ga d(x^k) + P_{\spanI^{\perp}}(v_j)}| \leq  \frac{\gamma}{2}(1+\sqrt{\alpha}) ~.
}
{~}\\

\noindent
As an orthogonal projector is a self-adjoint idempotent operator, we have for all $j\leq p$,
\eq{
|h[j]\ps{a_j}{P_{\spanI^{\perp}}(a_j)}|\leq \normi{h}\normd{P_{\spanI^{\perp}}(a_j)}^2,
}
where $\normd{P_{\spanI^{\perp}}(a_j)}^2$ is the squared $\ell_2$-norm of the projection of a Gaussian vector on the subspace $\spanI^\perp$ whose dimension is $n-k$. As $\spanI^{\perp}$ is independent of $a_j$, for $j \notin I$, $n \normd{P_{\spanI^{\perp}}(a_j)}^2$ follows a $\chi^2$ distribution with $n-k$ degrees of freedom. Using Lemma~\ref{LemmeBorneChi2} together with assumptions \eqref{eq-th-compress-2}-\eqref{eq-th-compress-gamma}, the following bound holds \wop
\eql{\label{compres2}
\max_{j\notin j} |h[j]\ps{a_j}{P_{\spanI^{\perp}}(a_j)}|\leq
2.5\normi{h}\leq \frac{\ga}{2}(1-\sqrt{\alpha})
}

In summary, \eqref{compres1} and \eqref{compres2} show that $(C_2)$ is
fulfilled with probability larger than $1-\frac{1}{3\sqrt{2\pi n}}
e^{-0.8n} -\frac{1}{3\sqrt{2\pi n}}e^{-0.3n} 
-\frac{1}{\sqrt{2\pi(n-k)}}e^{-0.009n}$.

\section{Proof of Partial Support Recovery}
\label{sec-PartialSupportRecovery}

To prove the first part of Theorem \ref{TheoPartialRecovery}, we need to show that with \wop, the extension  $x_1(\gamma)$ on $\mathbb{R}^p$ of the solution of 
\begin{equation}\label{eqlunpartial}
\min_{x\in\mathbb{R}^{|I|}}\frac{1}{2}\normd{y_1-\bA x}^2+\gamma\normu{x}
\end{equation}
with $y_1=P_{\bA}(y)$, is the solution of the Lasso. By definition, the support $J$ of this extension is included in $I$.

Proving this assertion amounts to showing that $x_1(\gamma)$ fulfills the necessary and sufficient optimality conditions
\begin{equation}\label{l1condF3}
	\choice{
		\transp{A_J}(y-Ax_1(\gamma))=\ga\sign{\ol{x_1(\gamma)}}, \\
		\forall l\notin J, \quad |\dotp{a_l}{y-Ax_1(\gamma)}| \leq \ga.
	}
\end{equation}   
Since $y_1=P_{\bA}(y)$ and $J \subset I$, 
$\transp{A_J}(y-Ax_1(\gamma))=\transp{A_J}(y_1-Ax_1(\gamma))$. In addition, as $x_1(\gamma)$ is the extension of the solution of \eqref{eqlunpartial}, the optimality conditions associated to \eqref{eqlunpartial} yield
\eq{
	\choice{
	\transp{A_J}(y-Ax_1(\gamma))=\ga\sign{\ol{x_1(\gamma)}}, \\
	\forall l\in (I\cap J^c), \quad |\dotp{a_l}{y-Ax_1(\gamma)}| \leq \ga.
	}
}
To complete the proof, it remains now to show that \wop
\begin{equation}\label{eqobj}
	\forall l\notin I, \quad
        |\dotp{a_l}{y-Ax_1(\gamma)}| \leq \ga.
\end{equation}
As in the proofs of Theorems \ref{TheoBruit} and \ref{TheoNotSparse2}, to bound these scalar products, the key argument is the independence between the vectors $(a_l)_{l\notin I}$ and the residual vector $y-Ax_1(\gamma)$.
{~}\\

We first need the following intermediate lemma.
\begin{lemme}\label{lemmedecroissance}
Let $A\in\mathbb{R}^{n\times k}$ such that $(\transp{A}A)$ is invertible. Take $x(\gamma)$ as a solution of the Lasso from observations $y\in \mathbb{R}^n$. The mapping $f: \mathbb{R}^{+*} \to \mathbb{R}^+$, $\gamma \mapsto f(\gamma)=\frac{\normd{y-Ax(\gamma)}}{\gamma}$ is well-defined and non-increasing.  
\end{lemme}

\begin{proof} 
The authors in \cite{osborne-homotopy} and \cite{dossal-topological} independently proved that under the assumptions of the lemma:
\begin{itemize}
\item the solution $x(\gamma)$ of the Lasso is unique;
\item there is a finite increasing sequence $(\gamma_t)_{t \leq K}$ with $\gamma_0=0$ and $\gamma_K=\normi{\transp{A}y}$ such that for all $t<K$, the sign and the support of $x(\gamma)$ are constant on each interval $(\gamma_{t},\gamma_{t+1})$.
\item $x(\gamma)$ is a continuous function of $\gamma$.
\end{itemize}
Moreover $x(\gamma)$ with support $J$ satisfies 
\begin{equation}\label{eqdefxgamma}
\ol{x(\gamma)}=A_J^+y-\gamma(\transp{A_J}A_J)^{-1}\sign{\ol{x(\gamma)}},
\end{equation} 
which implies that
\begin{equation*}
r(\gamma):=y-Ax(\gamma)=P_{A_J^{\perp}}(y)-\gamma A_J(\transp{A_J}A_J)^{-1}\sign{\ol{x(\gamma)}} ~.
\end{equation*}
Therefore, on each interval $(\gamma_{t},\gamma_{t+1})$, $r(\gamma)$ is an affine function of $\gamma$ which can be written 
\begin{equation*}
r(\gamma)=z-\gamma v,
\end{equation*}
where $z:=P_{A_J^{\perp}}(y)$ and $v:=A_J(\transp{A_J}A_J)^{-1}\sign{\ol{x(\gamma)}}$. As $v \in V_J$ and $z\in V_J^{\perp}$, the Pythagorean theorem allows to write for $\gamma\in(\gamma_t,\gamma_{t+1})$ that
\begin{equation}
\frac{\normd{r(\gamma)}^2}{\gamma^2}=\frac{\normd{z}^2}{\gamma^2}+\normd{v}^2 ~.
\end{equation}
We then deduce that $f(\gamma)=\frac{\normd{r(\gamma)}}{\gamma}$ is a non-increasing function of $\gamma$ on each interval $(\gamma_t,\gamma_{t+1})$. By continuity of $f$, it follows that $f$ is non-increasing on $\mathbb{R}^{+*}$.

\end{proof}

\begin{remarque}
If $(\bAt \bA)$ is not invertible, the Lasso may have several solutions. Nevertheless $r(\gamma)$ is always uniquely defined and the lemma should also apply. 
\end{remarque}

From Lemma~\ref{lemmedecroissance}, we deduce that $\frac{\normd{y_1-Ax_1(\gamma)}}{\gamma}$ is a non-increasing function of $\gamma$. Because $y_1\in \spanI$ and $\bA$ has full column-rank, we also have
\begin{equation*}
\lim_{\gamma \to 0}x_1(\gamma)=x_1,
\end{equation*}
where on $I$, the entries of $x_1$ are those of the unique vector of $\mathbb{R}^{|I|}$ such that $\bA x =y_1$. Therefore,
\eql{\label{eq-partial-x1x0}
x_1[i] = x_0[i] + (\bA^+ w)[i], \qforq i \in I ~.
}
Since $\bA$ is Gaussian and independent from $x_0$ and $w$, the support of $x_1$ is almost surely equal to $I$. Hence there exists $\gamma_1>0$ such that if $\gamma<\gamma_1$, the support and the sign of $x_1(\gamma)$ are equal to those of $x_1$. More precisely, if $\gamma<\gamma_1$, $x_1(\gamma)$ satisfies 
\begin{equation*}
\ol{x_1(\gamma)}=\ol{x_1}-\gamma(\bAt \bA)^{-1}\sign{\ol{x_1}} \qandq r(\gamma):=y_1-Ax_1(\gamma)=\gamma \bA(\bAt \bA)^{-1}\sign{\ol{x_1}} ~.
\end{equation*}  
It then follows that for $\gamma\in(0,\gamma_1)$,
\[
\frac{\normd{y_1-Ax_1(\gamma)}}{\gamma}=\normd{\bA(\bAt \bA)^{-1}\sign{\ol{x_1}}}.
\]
Now, since
\begin{equation*}
\normd{\bA(\bAt \bA)^{-1}\sign{\ol{x_1}}}^2=\dotp{(\bAt \bA)^{-1}\sign{\ol{x_1}}}{\sign{\ol{x_1}}},
\end{equation*}
we deduce that for all $\gamma>0$,
\[
\frac{\normd{y_1-Ax_1(\gamma)}}{\gamma}\leq \sqrt{|I|\rho((\bAt \bA)^{-1})},
\]
where $\rho((\bAt \bA)^{-1})$ is the spectral radius of $(\bAt \bA)^{-1}$. Using Lemma~\ref{LemmeVSWishart} with  $\beta<\left(1-\sqrt{\frac{k}{n}}\right)^2$ then leads to
\begin{equation}\label{eqborneresidu}
P\left(\frac{\normd{y_1-Ax_1(\gamma)}}{\gamma}\leq 
\sqrt{\frac{k}{\beta}}\right) \geq 1-e^{-\frac{n\left(1-\sqrt{\beta}-\sqrt{\frac{k}{n}}\right)^2}{2}} ~.
\end{equation}
By the Pythagorean theorem and the fact that $\normd{P_{\spanI^\perp} w} \leq \varepsilon$, we have 
\begin{eqnarray*}
\normd{y-Ax_1(\gamma)}^2 &=& \normd{y-y_1}^2 + \normd{y_1-Ax_1(\gamma)}^2 \\
			 &=& \norm{P_{\spanI^\perp} w}^2 + \normd{y_1-Ax_1(\gamma)}^2 \\
			 &\leq& \varepsilon^2+\normd{y_1-Ax_1(\gamma)}^2 ~.
\end{eqnarray*}
With similar arguments as those leading to \eqref{eq-exact-C2-max}, it can then be deduced that 
\begin{equation}
\max_{l\notin I}|\dotp{a_l}{y-Ax_1(\gamma)}|\leq
\sqrt{\frac{2\log p}{n}\left(\varepsilon^2+\frac{\gamma^2k}{\beta}\right)} ~.
\end{equation}
with probability larger than $1-e^{-\frac{n\left(1-\sqrt{\beta}-\sqrt{\frac{k}{n}}\right)^2}{2}}-\frac{1}{2\sqrt{\pi\log p}}$,
\\
If $k\leq \frac{\alpha \beta n}{2\log p}$ and $\gamma\geq\frac{\varepsilon}{\sqrt{1-\alpha}}\sqrt{\frac{2\log p}{n}}$, then $\sqrt{\frac{2\log p\left(\varepsilon^2+\frac{\gamma^2k}{\beta}\right)}{n}}\leq \gamma$, and therefore inequality \eqref{eqobj} is satisfied \wop. This ends the proof of the first part of the theorem.\\

Let's now turn to the proof of \eqref{eq-th-partial-l2}. 
To prove this inequality we notice that for large $\gamma$, the Lasso solution $x(\gamma)$ is also the extension of the solution of \eqref{eqlunpartial} \wop and we use the Lipschitz property of the mapping $\gamma \mapsto x_1(\gamma)$.

Indeed, by the triangle inequality, 
\begin{equation}
\normd{x_0-x_1(\gamma)}\leq \normd{x_0-x_1}+\normd{x_1-x_1(\gamma)} ~.
\end{equation}
Recalling from \eqref{eq-partial-x1x0} that $\ol{x_0}-\ol{x_1}=\bA^+ w$, it follows that  
\[
\normd{x_0-x_1}\leq \varepsilon \sqrt{\rho((\bAt \bA)^{-1})},
\]
which, using again Lemma~\ref{LemmeVSWishart}, leads to the bound
\[
\normd{x_0-x_1}\leq 2\varepsilon
\] 
with probability larger than $1-e^{-\frac{n}{2}\left(0.5-\sqrt\frac{k}{n}\right)^2}$. \\

For all $\gamma>0$, $x_1(\gamma)$ obeys \eqref{eqdefxgamma}, and since $\lim_{\gamma \to 0}x_1(\gamma)=x_1$, we get that 
\begin{equation}
\normd{x_1-x_1(\gamma)}\leq \gamma\max_{J \subset I,S\in\{-1,1\}^{|J|}} \normd{(\transp{A_J}A_J)^{-1}S}.
\end{equation} 
For all $J\subset I$, the inclusion principe tells us that $\rho((\transp{A_J}A_J)^{-1})\leq \rho((\bAt \bA)^{-1})$. Furthermore, for all $S\in \{-1,1\}^{|J|}$, $\normd{S}\leq \sqrt{k}$. Using Lemma~\ref{LemmeVSWishart} once again implies that
\begin{equation*}
P\left(\normd{x_1-x_1(\gamma)}\leq \gamma\sqrt{\frac{k}{\beta}}\right)\geq 1-e^{-\frac{n}{2}\left(1-\sqrt{\beta}-\sqrt{\frac{k}{n}}\right)^2}.
\end{equation*}
If $\gamma=\frac{\varepsilon}{\sqrt{1-\alpha}}\sqrt{\frac{2\log p}{n}}$ and $k\leq \frac{\alpha \beta n}{2\log p}$ , then \wop
\begin{equation*}
\normd{x_1-x_1(\gamma)}\leq \varepsilon\sqrt{\frac{\alpha}{1-\alpha}}~.
\end{equation*}
This concludes the proof.

\section{Numerical Illustrations}
\label{sec-numerics}

This section aims at providing empirical support of the sharpness of our bounds by assessing experimentally the quality of the constants involved in Theorem \ref{TheoBruit}. More specifically, we perform a probabilistic analysis of support and sign recovery, to show that the bounds \eqref{eq-sparsity-constr}, \eqref{eq-gamma-value} and \eqref{eq-minsnr-constr} are quite tight\footnote{The \textsc{Matlab} code to reproduce the figures are freely available for download from \url{http://www.ceremade.dauphine.fr/~peyre/codes/}.}.

In all the numerical tests, we use problems of size $(n,p) = (8000,32000)$ and $(n,p) = (3000,36000)$, corresponding to moderate and high redundancies. These are realistic high-dimensional settings in agreement with signal and image processing applications. We perform a randomized analysis, where the probability of exact recovery of supports and signs (sparsistency) are computed by Monte-Carlo sampling with respect to a probability distribution on the measurement matrix, $k$-sparse signals and on the noise $w$. As detailed in Section~\ref{subsec:setup}, the matrix $A$ is drawn from the Gaussian ensemble. We assume that the non-zero entries $x[i]$ for $i \in I(x)$ of a vector $x \in \RR^p$ are independent realizations of a Bernoulli variable taking equiprobable values $\{+T,-T\}$. We also assume that the noise $w$ is drawn from the uniform distribution on the sphere $\enscond{w \in \RR^n}{\norm{w}=\varepsilon}$. Since only the SNR matters in the bounds, we fix $\varepsilon = 1$ and only vary the value of $T$.

\newcommand{\twinfig}[2]{ %
	\begin{tabular}{@{}c@{\hspace{1mm}}c@{}}
	\includegraphics[width=0.48\linewidth]{#1}	&
	\includegraphics[width=0.48\linewidth]{#2} \\
	$(n,p) = (8000,32000)$ &
	$(n,p) = (3000,36000)$
	\end{tabular}
}

\myfigure{
	\twinfig{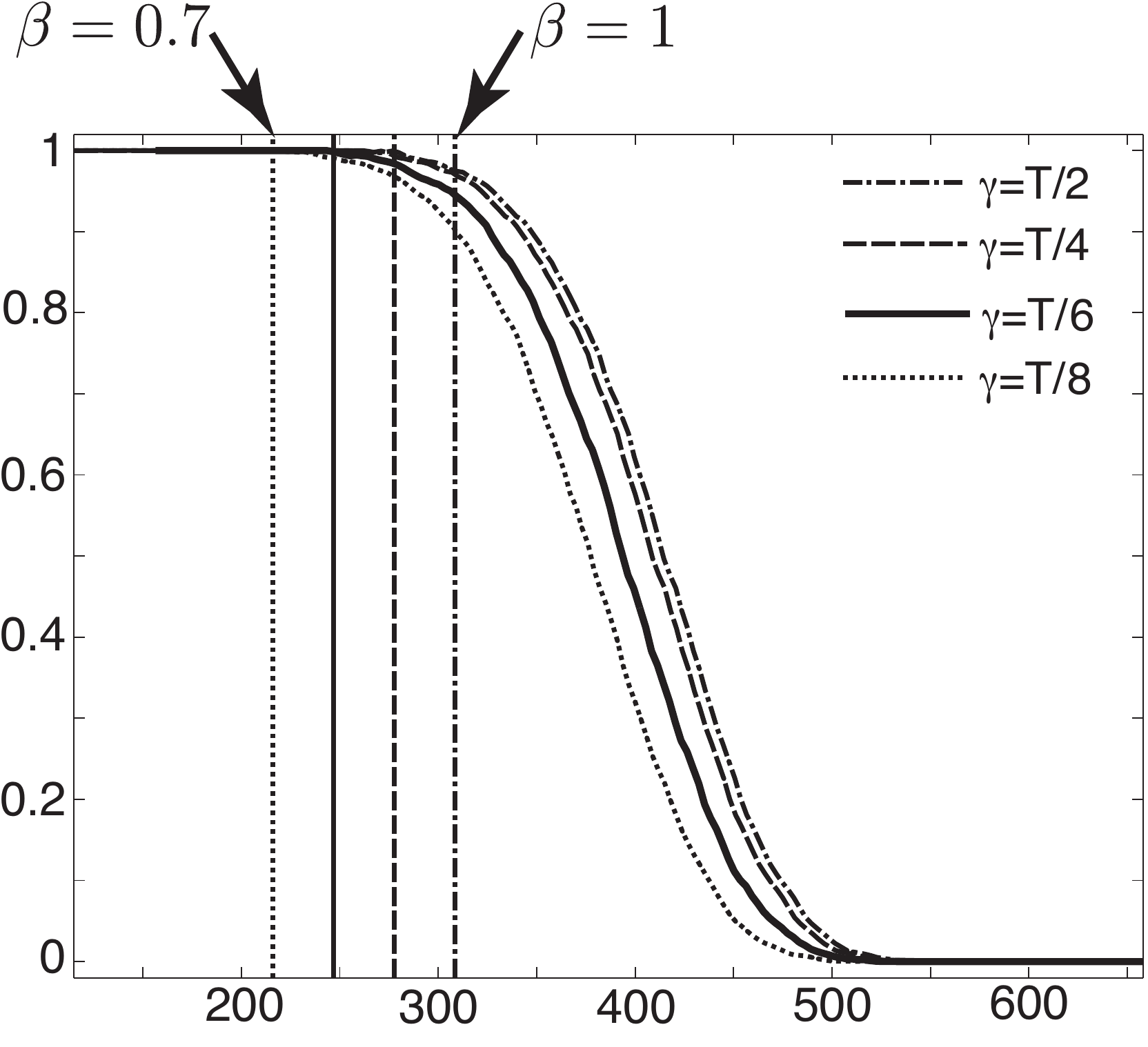}{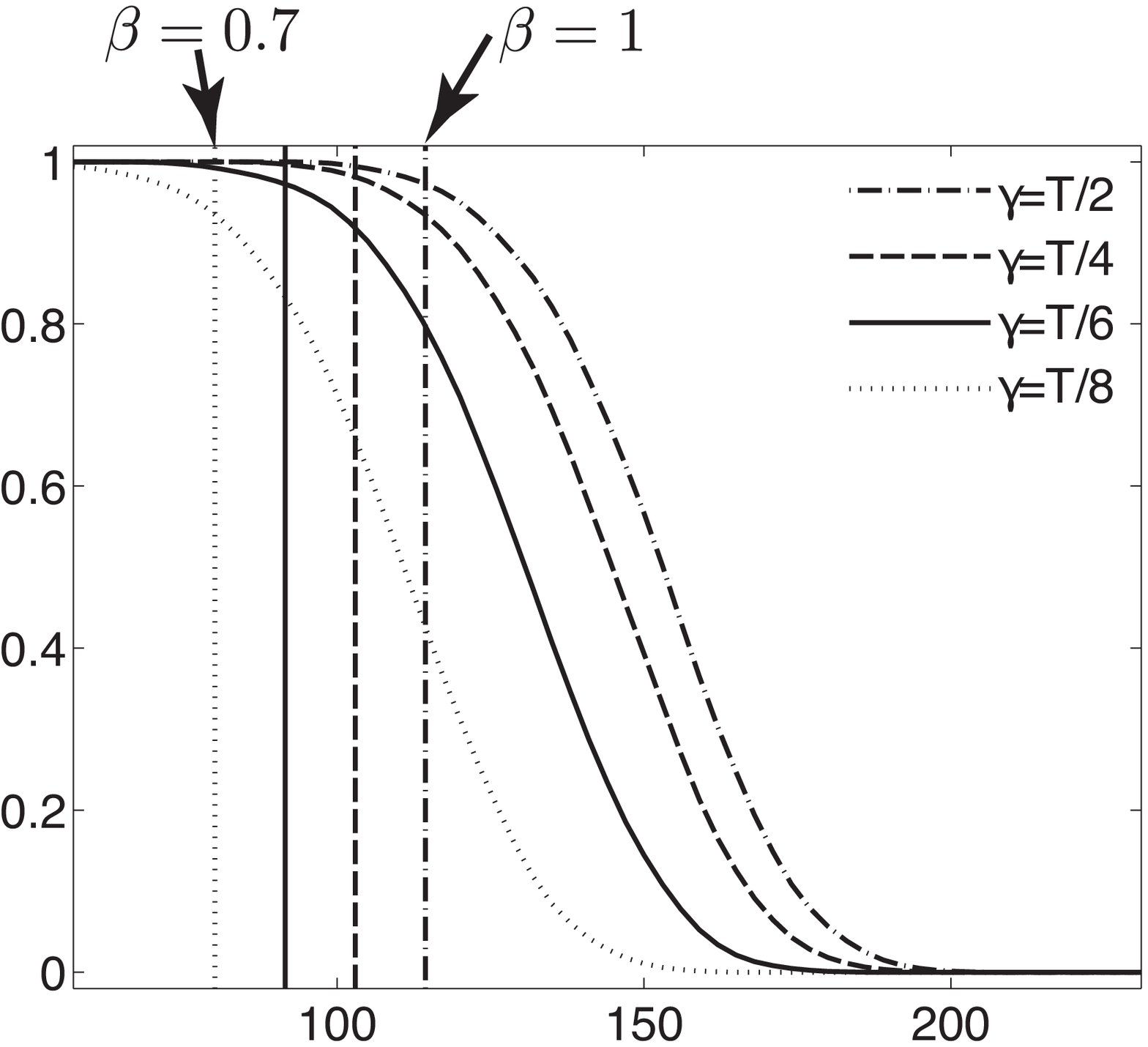}
}{ %
	Probability of sparsistency as a function of $k$ 
	and $\alpha = 0.8$.
	The vertical lines corresponds to our sparsistency bound $k_\beta$, from left to right, 
	for $\beta = 0.7, 0.8, 0.9, 1$.
}{fig-identifiability}

\paragraph{Challenging the sparsity bound~\eqref{eq-sparsity-constr}}

We first evaluate, for $\al=0.8$, and for a varying value of $k$, the probability of sparsistency given that 
\eql{
	T = \frac{\six\varepsilon}{\sqrt{1-\alpha}} \sqrt{\frac{2\log p}{n}}
	\qandq
	\gamma = \frac{T}{\six}
}
which are values in accordance with the bounds \eqref{eq-minsnr-constr} and \eqref{eq-gamma-value}.

In order to compute numerically this probability, for each $k$, we generate $1000$ sparse signals $x_0$ with $\normz{x_0}=k$, and check whether conditions $(C_1)$ and $(C_2)$ defined in Section~\ref{subsec_spars1} are satisfied. Figure~\ref{fig-identifiability} shows how this probability decays when $k$ increases. The vertical lines correspond to the critical sparsity thresholds
\eql{\label{eq-crit-sparsity}
	k_\beta = \frac{\alpha \beta n}{2\log p}
}
as identified by the bound \eqref{eq-sparsity-constr}. The estimated probability exhibits a typical phase transition that is located precisely around the critical value $k_\beta$ for $\beta$ close to one. This shows that our bound is quite sharp. We also display the same probability curve for other, less conservative, values of $\ga \in \{T/4, T/2\}$, which improves slightly the probability with respect to $\ga = T/\six$.

\paragraph{Challenging the regularization parameter value~\eqref{eq-gamma-value}}

We evaluate, for $(\al,\beta) = (0.8,0.8)$, the probability of sparsistency using a value of $\ga$ different from 
\eql{\label{eq-defn-gamma0}
	\ga_0 = \frac{\varepsilon}{\sqrt{1-\alpha}}\sqrt{\frac{2\log p}{n}}
}
given in \eqref{eq-gamma-value}, for which Theorem \ref{TheoBruit} is valid. 
We use the critical sparsity level $k=k_\beta$ defined in \eqref{eq-crit-sparsity}.
To study only the influence of $\ga$, we use a SNR that is infinite, meaning that $\varepsilon$ is negligible in comparison with $T$. This implies in particular that in this regime, only condition $(C_1)$ has to be checked to estimate the probability of sparsistency.

Figure~\ref{fig-gamma} shows the increase in this probability as the ratio $\ga/\ga_0$ increases. This makes sense because the signal is large with respect to the noise so that a large threshold should be preferred. One can see that at the critical value $\ga=\ga_0$ suggested by Theorem \ref{TheoBruit}, this probability is close to 1. This again confirms that the value \eqref{eq-gamma-value} of $\ga$ is quite sharp.

\myfigure{
	\twinfig{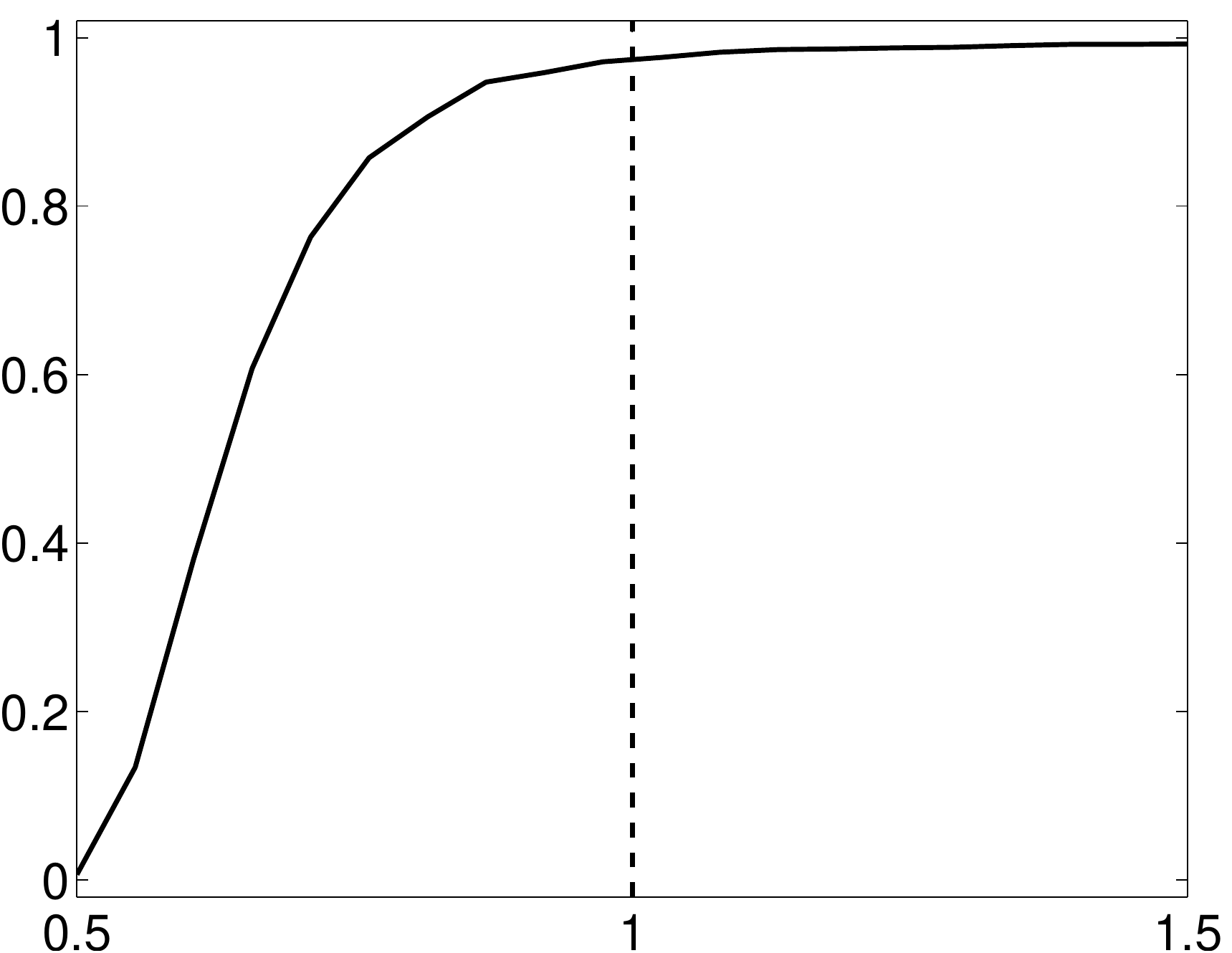}{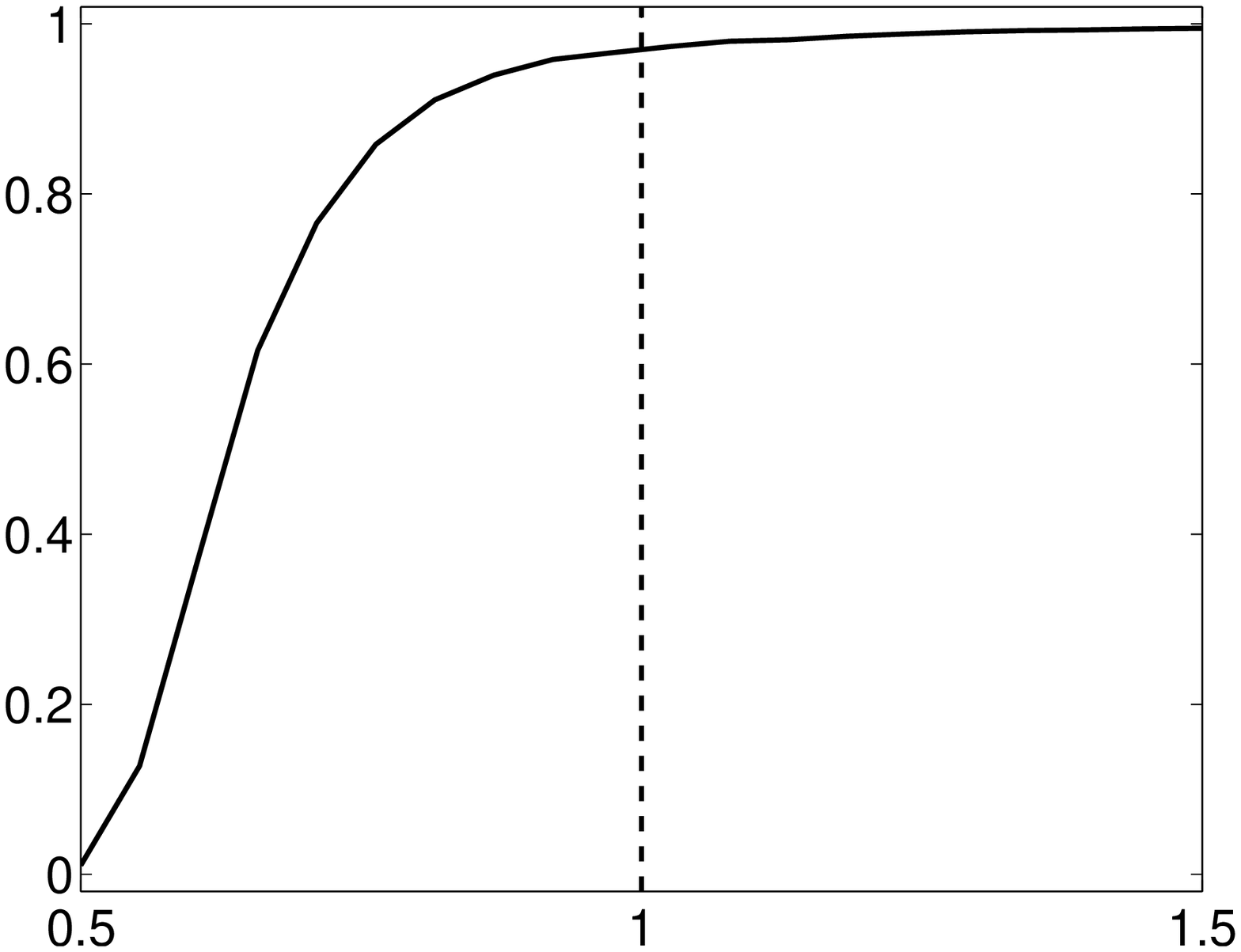}
}{ %
	Probability of support recovery for large $T$ as a function of $\ga/\ga_0$
	for $k=k_\beta$ and $(\alpha,\beta)=(0.8,0.8)$.
}{fig-gamma}

\paragraph{Challenging the signal-to-noise ratio \eqref{eq-minsnr-constr}}

Lastly, we estimate, for $(\al,\beta) = (0.8,0.8)$, the minimal signal level $T$ that is required to ensure the inclusion of the support, meaning that $I(x(\ga)) \subset I(x_0)$. We use the critical sparsity $k=k_\beta$ and $\ga=\ga_0$, with $k_\beta$ and $\ga_0$ as defined respetively in \eqref{eq-crit-sparsity} and \eqref{eq-defn-gamma0}. Since we are only interested in support inclusion, it is only needed to check condition $(C_2)$. 

The bound in \eqref{eq-minsnr-constr} suggests that $T \geq \six \ga_0$ is enough. Figure~\ref{fig-T} however shows that this bound is pessimistic, and that $T \geq 2 \ga_0$ appears to be enough to guarantee the support inclusion with high probability. A few reasons may explain this sub-optimality.
\begin{itemize}
\item There is no guarantee that the concentration lemmas we use are optimal.
\item The limit ratio $\frac{T}{\varepsilon}$ relies mainly on Lemma~\ref{LemmeBorneInf} and especially on the bound $1+4\sqrt{\betabis}$ in it. 
This bound can be improved by at least three ways.
\begin{itemize}
\item Using the same proof, the bound can be slightly enhanced by decaying the probability of success. 
\item The result in the lemma is non-asymptotic. The bound and the probability were computed to be available for all $\alpha\leq 1,\beta\leq 1$ and for all $p\geq 1212$. With the values used in the numerical experiments, and decaying a bit the probability of success, the bound can turn into $1+2.7\sqrt{\betabis}$, yielding a better bound $T\geq 4.37 \ga_0$.
\item In the proof of Lemma~\ref{LemmeBorneInf}, the inequality $\norm{B_i}_2\leq \rho(B)$, is used, where $\rho(B)$ is the spectral radius of $B$. This bound is available for any matrix, but one might perhaps do better by exploiting Gaussiannity of the measurement matrix.
\end{itemize}   
\end{itemize}

\myfigure{
	\twinfig{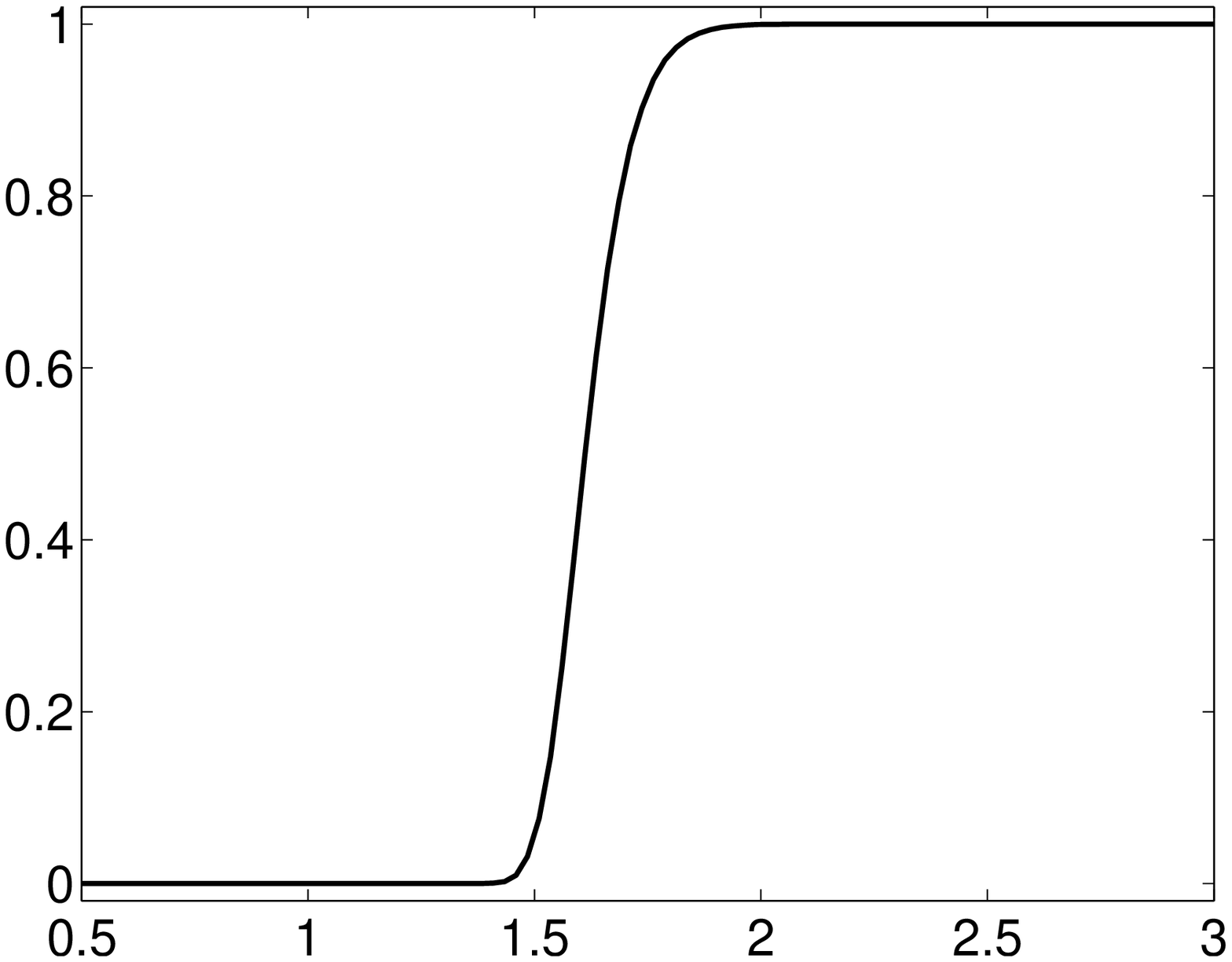}{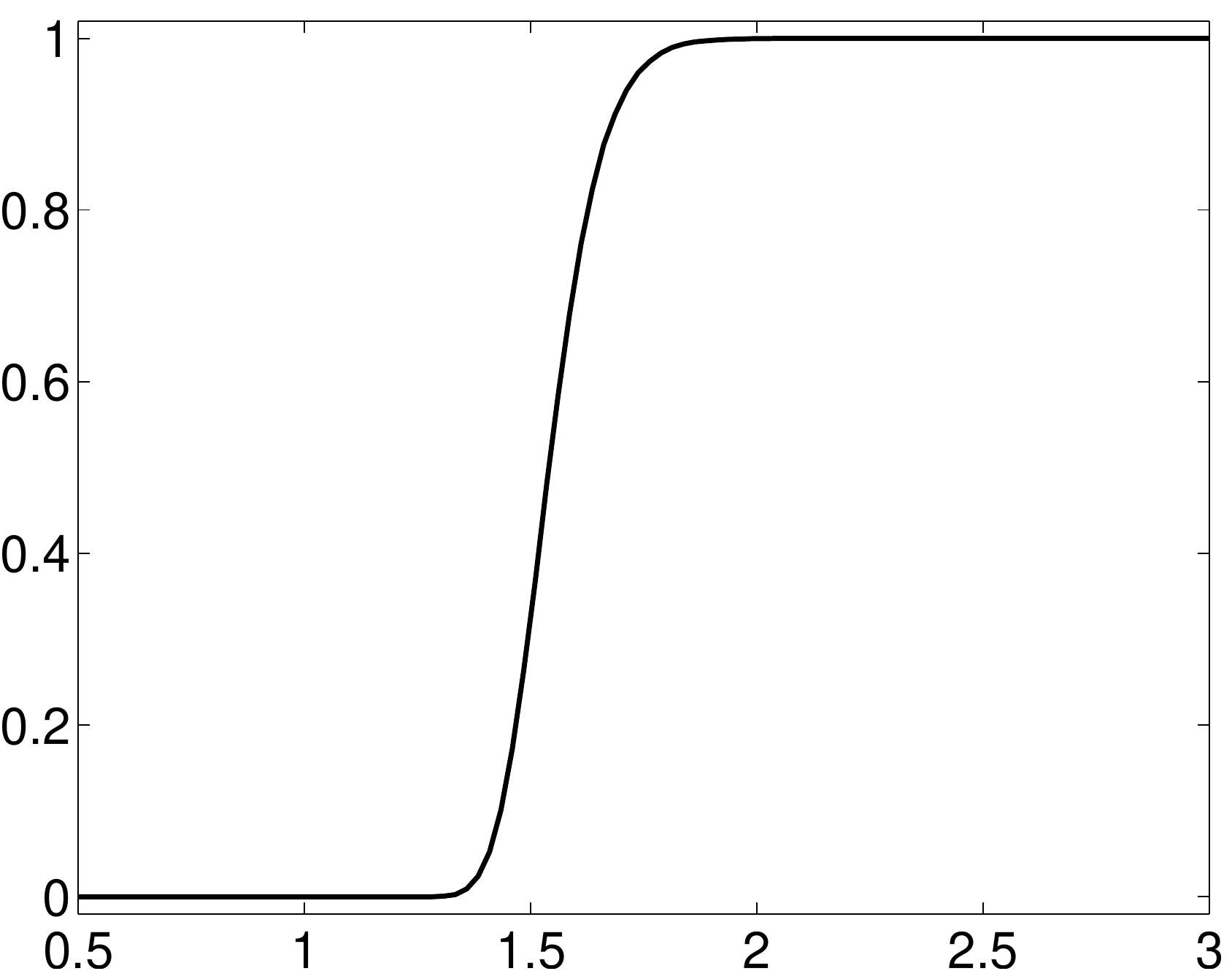}
}{ %
	Probability of support inclusion as a function of $T/\ga_0$
	for $k=k_\beta$ and $(\alpha,\beta)=(0.8,0.8)$.
}{fig-T}

\section*{Conclusion}
\label{sec-conclusion}

This paper has presented a novel analysis of the sparsistency of the Lasso from noisy Gaussian measurements. We derived sharp bounds on the sparsity of the signal to guarantee sparsistency with high probability. This result is extended to handle compressible signals and to establish sharp $\ldeux$-consistency. A distinctive feature of our analysis is that it provides explicit constants for the three key parameters of the problem: the sparsity of the signal, the minimal signal-to-noise ratio and the Lasso regularization parameter. Numerical results support the claim that these constants are either sharp or at least reasonably well behaved.



\appendix
\section{Properties of Wishart Matrices}
\label{subsec-lemma-spectral}

\subsection{Signs of non-diagonal entries of an inverse Wishart matrix}
\label{subsec-lemma-spectral-inverse}

\begin{lemme}\label{LemmeWishartRademacher}
If $B\in \RR^{k\times k} $ is the inverse of a Wishart matrix, then
for all $i\leq k$, the variables $(\sign{B_{i,j}}, j\neq i)$ form
a Rademacher sequence, that is they are independent and uniformly
distributed on $\{-1,1\}$. Moreover this sequence is independent of
$B_{i,i}$ , and of $(|B_{i,j}|)_{j\neq i}$.

\end{lemme}
\begin{proof}
If $B=(B_{i,j})_{i\leq
 k,j\leq k} \in \RR^{k\times k} $ is the inverse of a Wishart matrix, then
$B=(\transp{A}A)^{-1}$ where $A  \in \mathcal{M}_{n,k}(\mathbb{R})$ is a
Gaussian matrix. Let $E \in \mathcal{M}_{k,k}(\mathbb{R})$ be diagonal such that for all $1\leq
i\leq k, |E_{i,i}|=1$. 
Then $\transp{(AE)}AE=E\transp{A}AE$, hence  $(\transp{(AE)}(AE))^{-1}=E(\transp{A}A)^{-1}E$. 
Therefore the entries of $C=(\transp{(AE)}AE)^{-1}$ are
$C_{i,j} = E_{i,i}E_{j,j}B_{i,j}$ for $1\leq i,j\leq k$.\\
But $A$ and $AE$ have the same law, hence $B$ and $C$ also have the same law.
Hence for all $(\epsilon_j)_{j\leq k, j\neq i} \in
\{-1,1\}^{k-1}$, the laws of $(B_{i,1}, \ldots, B_{i,k})$ and
$(\epsilon_1 B_{i,1},
\ldots, B_{i,i}, \ldots, \epsilon_k B_{i,k})$ are the same. This implies that
the variables $(\sign{B_{i,j}}, j\neq i)$ form a Rademacher sequence, and  this sequence is independent of $B_{i,i}$, and of $(|B_{i,j}|)_{j\neq i}$.
\end{proof}

\subsection{Extreme eigenvalues of a Wishart matrix}
\label{subsec-lemma-spectral-extreme}

The proof of the following lemma can be found in \cite[page 42]{davidson-book}.
\begin{lemme}\label{LemmeVSWishart}
	If $A \in  \RR^{n \times k}$ is a Gaussian matrix whose
        coefficients are centered of variance $\frac{1}{n}$, then the
        maximal and minimal eigenvalues  of the Wishart matrix
        $B=\transp{A} A$ satisfy for all $t>0$
	\eq{ 
		P\left(\lambda_{\max}(B)\geq \left(1+\sqrt{\frac{k}{n}}+t\right)^2\right)\leq e^{-\frac{nt^2}{2}}
	}
	and 
	\eq{
		P\left(\lambda_{\min}(B)\leq \left(1-\sqrt{\frac{k}{n}}-t\right)^2\right)\leq e^{-\frac{nt^2}{2}}
	}
\end{lemme}

\subsection{Sup-norm of a projected Rademacher sequence}
\label{subsec-lemma-spectral-sup}

\begin{lemme}\label{LemmeBorneInf}
	If $C \in \RR^{n \times k}$ is a Gaussian matrix, with $k\leq
        \frac{n\betabis}{2 \log p}$ with $0<\betabis\leq 1$  
	and if $S \in \{-1,1\}^{k}$ is drawn independently from $C$,
	then if $p \geq 1212$, 
	\begin{equation*}
		P\left(\normi{(\transp{C}C)^{-1}S} \leq
                  1+4\sqrt{\betabis}\right) \geq 1-kp^{-1.28}-2e^{-\frac{n\betabis(0.75\sqrt{2}-1)^2}{4\log p}} ~.
	\end{equation*}
\end{lemme}

\begin{proof}
	We use the following splitting
	\eq{
		(\transp{C}C)^{-1}=I+((\transp{C}C)^{-1}-I)=I+B.
	} 
	This shows that  
	\eq{
		\normi{(\transp{C}C)^{-1}S} \leq \normi{S}
		+ \normi{BS}=1+\normi{BS}.
	}
	
One can then observe that  $(BS)[i]=\sum_{j\leq k}|B_{i,j}|S[j]\sign{B_{i,j}}$; one has $B_{i,i}>0$, and according to Lemma~\ref{LemmeWishartRademacher}, for given $i$, the variables $\sign{B_{i,j}}_{j\neq i}$ form a Rademacher sequence (this means that they are independent and uniformly distributed on $\{-1,1\}$), and this sequence is independent of $B_{i,i}$ and of $(|B_{i,j}|)_{j\neq i}$. Hence one can apply Hoeffding's Lemma~\ref{LemmeHoeffding} (multiplying the line by an independent variable uniform on $\{-1,1\}$ to take care of the fact that $\sign{B_{i,i}}$ is not uniformly distributed), thus getting for any $i\leq k$ and any $t>0$,

\begin{equation}\label{eqBi}
	P\left(\left|\sum_{j=1}^k B_{i,j}S[j]\right|\geq t\normd{B_i}\right)\leq e^{-\frac{t^2}{2}} ~.
\end{equation}

Now, for all  $i\leq k$, $\normd{B_i}\leq \rho(B)$, where $\rho(B)$ is the spectral radius of $B$. Using Lemma~\ref{LemmeVSWishart} with $t=(0.75-\frac{1}{\sqrt{2}})\sqrt{\frac{\betabis}{\log p}}$ and the fact that $\frac{k}{n}\leq \frac{\betabis}{2\log p}$, we get
\eq{
	P\left(\lambda_{\min}(\transp{C}C) \leq 
	\left(1-0.75\sqrt{\frac{\betabis}{\log p}}\right)^2\right)\leq e^{-\frac{n\betabis(0.75\sqrt{2}-1)^2)}{4\log p}} ~.
}
Consequently
\eq{
	P\left(\lambda_{\max}((\transp{C}C)^{-1}) \geq 
	\left(1-0.75\sqrt{\frac{\betabis}{\log p}}\right)^{-2}\right)\leq
      e^{-\frac{(0.75\sqrt{2}-1)^2\betabis n}{4\log p}} ~.
}
Similarly, we have
\eq{
	P\left(\lambda_{\min}((\transp{C}C)^{-1})\leq 
	\left(1+0.75\sqrt{\frac{\betabis}{\log p}}\right)^{-2}\right)\leq
      e^{-\frac{(0.75\sqrt{2}-1)^2 \betabis n}{4\log p}} ~.
}
It finally follows that with probability larger than $1-2e^{-\frac{n\betabis(0.75\sqrt{2}-1)^2)}{4\log p}}$,
\eq{
	\rho(B)\leq \max\left(\left|(1+0.75\sqrt{\frac{\betabis}{\log p}})^{-2}-1\right|,
	\left|(1-0.75\sqrt{\frac{\betabis}{\log p}})^{-2}-1\right| \right) ~.
}
In particular, taking $\frac{\log(p)}{\betabis} \geq \frac{15^2}{(17-\sqrt{129})^2} \simeq 7.07 $ leads to $\rho(B)\leq
2.5\sqrt{\frac{\betabis}{\log p}}$ with probability greater than $1-2e^{-\frac{n\betabis(0.75\sqrt{2}-1)^2}{4\log p}}$.

Using this bound in \eqref{eqBi} with $t=1.6\sqrt{\log(p)}$ 
yields  

\begin{align*}
 	P\left(\normi{BS}\geq 4\sqrt{\betabis} \right)&\leq
        P\left(\normi{BS}\geq  t\normd{B_i} 
          \mbox{ and }
        \rho(B) \leq 2.5\sqrt{\frac{\betabis}{\log p}}\right)\\
      & +
      P\left(\rho(B)\geq 2.5\sqrt{\frac{\betabis}{\log p}}\right)\\
      &\leq
      kp^{-1.28} + 2e^{-\frac{n\betabis(0.75\sqrt{2}-1)^2}{4\log p}} ~.
 \end{align*}

If we set  $\frac{\log(p)}{\betabis} \geq 7.08 $, the following holds,
\eq{
	P\left(\normi{(\transp{C}C)^{-1}S} \leq 1+4\sqrt{\betabis} \right)
	\geq 1-kp^{-1.28}-2e^{-\frac{n\betabis(0.75\sqrt{2}-1)^2}{4\log p}} ~.
}

\end{proof}

\begin{remarque}
It is worth noting that if $\frac{\log p}{\betabis}\geq
16.2$ as in the numerical experiments ($\betabis= 0.64, p=32000$),
one can adapt this proof and, by loosing a bit on the probability
(i.e. applying the concentration lemmas with smaller values of $t$), one
can get $\normi{(\transp{C}C)^{-1}S} \leq 1+2.7\sqrt{\betabis}$ \wop.
\end{remarque}
\subsection{Rotation invariance}
\label{subsec-rotinv}

\begin{lemme}\label{lem-rotinv}
If  $C \in \RR^{n \times k}$ is a Gaussian matrix, and $w \in \RR^n$ is independent of $C$, the law of $C^+ w$ is invariant under orthogonal transforms on $\RR^k$. 
\end{lemme}
\begin{proof}
	If  $C \in \RR^{n \times k}$ is a Gaussian matrix, then for
        any orthogonal matrix $U \in \RR^{k \times k}$, $D=CU$  and
        $C$ have the same distribution. The law of $D^+w$ and $C^+w$ are thus the same.
	Since for all $w$, one has
	\eq{
		D^+ w = U^{-1} C^+ w,
	} 
	the law of $U^{-1} C^+ w$ is the same as that 
	of $C^+ w$. 
\end{proof}

\subsection{Distribution of a quadratic form}
\label{subsec-lemma-spectral-quadratic}

The following lemma is a consequence of \cite[Theorem 3.2.12]{muirhead-book}.
\begin{lemme}\label{LemmeMuirhead}
If $B$ is a Wishart matrix as described in Lemma~\ref{LemmeVSWishart}, then for all $X \in \RR^k $ independent of $B$, the random variable $\frac{n\normd{X}^2} {\transp{X}B^{-1}X}$ follows a $\chi^2$ distribution with $n-k+1$ degrees of freedom.
\end{lemme}

\section{Concentration inequalities}
\label{subsec-lemma-concentration}

The following lemma is well known; a proof can be found in \cite{matousek-book}.
\begin{lemme}\label{lem-unifsphere}
Let $\mu_k$ denote the uniform probability on the unit sphere
$\Sphere^{k-1}$  in $\RR^k$, and let $A\subset \Sphere^{k-1}$ such that
$\mu_k(A)\geq \frac{1}{2}$. Then $\mu_k( \{x\in \Sphere^{k-1},
d(x,A)\leq \epsilon\}) \geq
1-2e^{-\frac{k\epsilon^2}{2}}$.
As a corollary, $\mu_k(x\in \Sphere^{k-1}, |x_1|\leq \epsilon\} \geq 1-4e^{-\frac{k\epsilon^2}{2}}$.
\end{lemme}

The following lemma is due to Cai et Silverman, see \cite{CaiSilverman01}.
\begin{lemme}\label{LemmeBorneChi2}
If $X$ follows a $\chi^2$ distribution with $k$ degrees of freedom,
then for all $\delta>0$,

\eq{
	P\left(X>(1+\delta)k\right)\leq \frac{1}{\sqrt{2\pi k}\delta}e^{-\frac{k}{2}(\delta- \log (1+\delta))}
}

\end{lemme}

The following lemma is due to Hoeffding, see \cite{Hoeffding63}.
\begin{lemme}\label{LemmeBorneChi2bis}
If $X$ follows a $\chi^2$ distribution with $k$ degrees of freedom,
then for all $\delta>0$,

\eq{
	P(X<(1-\delta)k)\leq e^{\frac{k\log(1-\delta)}{2}}
}
\end{lemme}

\gab{Precise this.}
The following lemma can be obtained by applying the Chernoff-Hoeffding inequality.
\begin{lemme}\label{LemmeHoeffding}
	If $(\varepsilon_i)_{i\leq k}$ is a Rademacher sequence,
        then for all  $a=(a_i)_{i\leq k}\in \RR^k$ and for all $t>0$,
	\eq{	
		P\left(\left|\sum_{i=1}^k \varepsilon_i a_i\right|\geq t\normd{a}\right)\leq e^{-\frac{t^2}{2}} ~.
	}
\end{lemme}

\bibliographystyle{elsarticle-num}
\bibliography{bibliography2}

\begin{thebibliography}{10}
\expandafter\ifx\csname url\endcsname\relax
  \def\url#1{\texttt{#1}}\fi
\expandafter\ifx\csname urlprefix\endcsname\relax\def\urlprefix{URL }\fi
\expandafter\ifx\csname href\endcsname\relax
  \def\href#1#2{#2} \def\path#1{#1}\fi

\bibitem{candes-robust}
E.~Cand\`{e}s, J.~Romberg, T.~Tao, Robust uncertainty principles: {E}xact
  signal reconstruction from highly incomplete frequency information, IEEE
  Trans. Info. Theory 52~(2) (2006) 489--509.

\bibitem{candes-near-optimal}
E.~Cand\`{e}s, T.~Tao, Near-optimal signal recovery from random projections:
  Universal encoding strategies?, IEEE Trans. Info. Theory 52~(12) (2006)
  5406--5425.

\bibitem{donoho-cs}
D.~Donoho, Compressed sensing, IEEE Trans. Info. Theory 52~(4) (2006)
  1289--1306.

\bibitem{chen-basis-pursuit}
S.~S. Chen, D.~Donoho, M.~Saunders, Atomic decomposition by basis pursuit, SIAM
  Journal on Scientific Computing 20~(1) (1998) 33--61.

\bibitem{TibshiraniLasso96}
R.~Tibshirani, Regression shrinkage and selection via the {Lasso}, Journal of
  the Royal Statistical Society 58~(1) (1996) 267--288.

\bibitem{candes-dantzig}
E.~J. Cand\`{e}s, T.~Tao, Rejoinder: the {D}antzig selector: statistical
  estimation when $p$ is much larger than $n$, Annals of Statistics 35~(6)
  (2007) 2392--2404.

\bibitem{BickelLassoDantzig07}
P.~J. Bickel, Y.~Ritov, A.~Tsybakov, Simultaneous analysis of lasso and
  {D}antzig selector, Annals of Statistics 37 (2009) 1705--1732.

\bibitem{osborne-homotopy}
M.~R. Osborne, B.~Presnell, B.~A. Turlach, On the lasso and its dual, Journal
  of Computational and Graphical Statistics 9~(2) (2000) 319--337.

\bibitem{EfronLars}
B.~Efron, T.~Hastie, I.~Johnstone, R.~Tibshirani, Least angle regression,
  Annals of Statistics 32~(2) (2004) 407--499.

\bibitem{donoho-homotopy}
D.~L. Donoho, Y.~Tsaig, Fast solution of $\ell_1$-norm minimization problems
  when the solution may be sparse, IEEE Trans. Info. Theory 54~(11) (2008)
  4789--4812.

\bibitem{figueiredo-nowak-em}
M.~Figueiredo, R.~Nowak, {An {EM} Algorithm for Wavelet-Based Image
  Restoration}, IEEE Trans. Image Proc. 12~(8) (2003) 906--916.

\bibitem{daubechies-iterated}
I.~Daubechies, M.~Defrise, C.~D. Mol, An iterative thresholding algorithm for
  linear inverse problems with a sparsity constraint, Commun. on Pure and Appl.
  Math. 57 (2004) 1413--1541.

\bibitem{bect-chambolle-iterative}
J.~Bect, L.~Blanc~F\'eraud, G.~Aubert, A.~Chambolle, A $\ell_1$-unified
  variational framework for image restoration, in: Proc. of ECCV04,
  Springer-Verlag, 2004, pp. Vol IV: 1--13.

\bibitem{combettes-proximal}
P.~L. Combettes, V.~R. Wajs, Signal recovery by proximal forward-backward
  splitting, SIAM Journal on Multiscale Modeling and Simulation 4~(4) (2005)
  1168--1200.

\bibitem{figueiredo-grad-projection}
M.~A.~T. Figueiredo, R.~D. Nowak, S.~J. Wright, Gradient projection for sparse
  reconstruction: Application to compressed sensing and other inverse problems,
  IEEE Journal of Selected Topics in Signal Processing 1~(4) (2007) 586--598.

\bibitem{bioucas-twist}
J.~M.~B. Dias, M.~A.~T. Figueiredo, A new tw{IST}: Two-step iterative
  shrinkage/thresholding algorithms for image restoration, IEEE Trans. Image
  Proc. 16~(12) (2007) 2992--3004.

\bibitem{nesterov-gradient}
Y.~Nesterov, Gradient methods for minimizing composite objective function, CORE
  Discussion Papers 2007076, Universit{\'e} catholique de Louvain, Center for
  Operations Research and Econometrics (CORE) (Sep. 2007).

\bibitem{beck-fista}
A.~Beck, M.~Teboulle, A fast iterative shrinkage-thresholding algorithm for
  linear inverse problems, Journal on Imaging Sciences 2~(1) (2009) 183--202.

\bibitem{combettes-dr}
P.~L. Combettes, J.-C. Pesquet, {A Douglas-Rachford splitting approach to
  nonsmooth convex variational signal recovery}, IEEE Journal of Selected
  Topics in Signal Processing 1~(4) (2007) 564--574.

\bibitem{Fadili09}
M.~Fadili, J.-L. Starck, Monotone operator splitting for fast sparse solutions
  of inverse problems, in: Proc. of IEEE ICIP, Cairo, Egypt, 2009.

\bibitem{StarckFadiliBook10}
J.-L. Starck, F.~Murtagh, M.~Fadili, Sparse Signal and Image Processing:
  Wavelets, Curvelets and Morphological Diversity, Cambridge University Press,
  Cambridge, UK, 2010.

\bibitem{CandesPlan09}
E.~J. {Cand{\`e}s}, Y.~{Plan}, {Near-ideal model selection by $\ell_1$
  minimization}, Annals of Statistics 37~(5A) (2009) 2145--2177.

\bibitem{donoho-stable-recovery}
D.~L. Donoho, M.~Elad, V.~N. Temlyakov, Stable recovery of sparse overcomplete
  representations in the presence of noise, IEEE Trans. Info. Theory 52~(1)
  (2006) 6--18.

\bibitem{Meinshausen06}
N.~Meinshausen, P.~B\"uhlmann, High-dimensional graphs and variable selection
  with the lasso, Ann. Statist. 34~(3) (2006) 1436--1462.

\bibitem{Greenshtein06}
E.~Greenshtein, Best subset selection, persistence in high-dimensional
  statistical learning and optimization under $\ell_1$ constraint, Annals of
  Statistics 34 (2006) 2367--2386.

\bibitem{tropp-just-relax}
J.~A. Tropp, Just relax: convex programming methods for identifying sparse
  signals in noise, IEEE Trans. Info. Theory 52~(3) (2006) 1030--1051.

\bibitem{wainwright-sharp-thresh}
M.~J. Wainwright, Sharp thresholds for high-dimensional and noisy sparsity
  recovery using $\ell_1$-constrained quadratic programming (lasso), IEEE
  Trans. Info. Theory 55~(5) (2009) 2183--2202.

\bibitem{ZhaoYu06}
P.~Zhao, B.~Yu, On model selection consistency of lasso, J. Mach. Learn. Res. 7
  (2006) 2541--2563.

\bibitem{Zou06}
H.~Zou, The adaptive lasso and its oracle properties, Journal of the American
  Statistical Association 101~(476) (2006) 1418--1429.

\bibitem{fuchs-bounded-noise}
J.~Fuchs, {Recovery of exact sparse representations in the presence of bounded
  noise}, IEEE Trans. Info. Theory 51~(10) (2005) 3601--3608.

\bibitem{Bunea08}
F.~Bunea, Consistent selection via the lasso for high dimensional approximating
  regression models, in: Pushing the Limits of Contemporary Statistics:
  Contributions in Honor of Jayanta K. Ghosh, Vol.~3, Institute of Mathematical
  Statistics, 2008, pp. 122--137.

\bibitem{zhou-privacy}
S.~Zhou, J.~D. Lafferty, L.~A. Wasserman, Compressed and privacy-sensitive
  sparse regression, IEEE Trans. Info. Theory 55~(2) (2009) 846--866.

\bibitem{fuchs-redundant-bases}
J.-J. Fuchs, On sparse representations in arbitrary redundant bases, IEEE
  Trans. Info. Theory 50~(6) (2004) 1341--1344.

\bibitem{Meinshausen09}
N.~Meinshausen, B.~Yu, Lasso-type recovery of sparse representations for
  high-dimensional data, Ann. Statist. 37~(1) (2009) 246---270.

\bibitem{vandeGeer09}
S.~A. van~de Geer, P.~B\"{u}hlmann, On the conditions used to prove oracle
  results for the lasso, Electron. J. Statist. 3 (2009) 1360--1392.

\bibitem{Bach08}
F.~R. Bach, Consistency of the group lasso and multiple kernel learning, J.
  Mach. Learn. Res. 9 (2008) 1179--1225.

\bibitem{nardi-lasso-asymp}
Y.~Nardi, A.~Rinaldo, On the asymptotic properties of the group lasso estimator
  for linear models, Electron. J. Statist. 2 (2008) 605--633.

\bibitem{YuanLin06}
M.~Yuan, Y.~Lin, {Model selection and estimation in regression with grouped
  variables}, Journal of the Royal Statistical Society: Series B (Statistical
  Methodology) 68~(1) (2006) 49--67.

\bibitem{tropp-NormsRandom}
J.~A. Tropp, Norms of random submatrices and sparse approximation, C. R. Math.
  Acad. Sci. 346 (2008) 1271--1274.

\bibitem{omidiran-subset}
D.~Omidiran, M.~J. Wainwright, High-dimensional subset recovery in noise:
  Sparsified measurements without loss of statistical efficiency, Tech. Rep.
  753, UC Berkeley (2008).

\bibitem{candes-reweighted-l1}
E.~J. Candes, M.~B. Wakin, S.~P. Boyd, Enhancing sparsity by reweighted {L1}
  minimization, J. Fourier Anal. Appl. 14~(5) (2008) 877--905.

\bibitem{Huang06}
J.~Huang, S.~Ma, C.-H. Zhang, Adaptive lasso for sparse high dimensional
  regression models, Tech. rep., Univ. of Iowa (2006).

\bibitem{fan-overview-selection}
J.~Fan, J.~Lv, A selective overview of variable selection in high dimensional
  feature space (invited review article), To appear in Statistica Sinica.

\bibitem{Zhang09}
T.~Zhang, Some sharp performance bounds for least squares regression with
  $\ell_l1$ regularization, Annals of Statistics 37 (2009) 2109--2144.

\bibitem{Wasserman09}
L.~Wasserman, K.~Roeder, High dimensional variable selection, Annals of
  statistics 37 (2009) 2178--2201.

\bibitem{geer-thesh-adap-lasso}
S.~A. van~de Geer, P.~B\"{u}hlmann, S.~Zhou, Prediction and variable selection
  with the adaptive lasso, Tech. Rep. arXiv:1001.5176v2 (2010).

\bibitem{wainwright-info-limits}
M.~J. Wainwright, Information-theoretic limits on sparsity recovery in the
  high-dimensional and noisy setting, IEEE Trans. Info. Theory 55~(12) (2009)
  5728--5741.

\bibitem{fletcher-sparse-pattern}
A.~K. Fletcher, S.~Rangan, V.~K. Goyal, Necessary and sufficient conditions on
  sparsity pattern recovery, {IEEE} Trans. on Information Theory 55~(12) (2009)
  5758--5772.

\bibitem{akcakaya-shannon}
M.~Ak{\c c}akaya, V.~Tarokh, Shannon theoretic limits on noisy compressive
  sampling, {IEEE} Trans. on Information Theory 56~(1) (2010) 492--504.

\bibitem{Reeves08}
G.~Reeves, M.~Gastpar, Sampling bounds for sparse support recovery in the
  presence of noise, in: Proceedings IEEE Int. Symp. on Inform. Theory, 2008,
  pp. 2187--2191.

\bibitem{Wang2010}
W.~Wang, M.~J. Wainwright, K.~Ramchandran, Information-theoretic limits on
  sparse support recovery: Dense versus sparse measurements, {IEEE} Trans. on
  Information Theory 56~(6) (2010) 2967--2979.

\bibitem{Aeron2010}
S.~Aeron, V.~Saligrama, M.~Zhao, Information theoretic bounds for compressed
  sensing, {IEEE} Trans. on Information Theory 56~(10) (2010) 5111--5130.

\bibitem{Saligrama2010}
V.~Saligrama, M.~Zhao, Thresholded basis pursuit: An lp algorithm for achieving
  optimal support recovery for sparse and approximately sparse signals from
  noisy random measurements, Tech. Rep. arxiv 0809.4883v3 (2010).

\bibitem{Reeves10}
G.~Reeves, M.~Gastpar, Approximate sparsity pattern recovery:
  Information-theoretic lower bounds, Tech. Rep. arXiv:1002.4458v1 (2010).

\bibitem{hormati-estimation}
A.~Hormati, A.~Karbasi, S.~Mohajer, M.~Vetterli, An estimation theoretic
  approach for sparsity pattern recovery in the noisy setting, Tech. Rep.
  LCAV-ARTICLE-2009-014, EPFL (2009).

\bibitem{Tune09}
P.~Tune, S.~R. Bhaskaran, S.~Hanly, Number of measurements in sparse signal
  recovery, in: Proceedings IEEE Int. Symp. on Inform. Theory, 2009, pp.
  16--20.

\bibitem{rad-sharp-pattern}
K.~R. Rad, Sharp sufficient conditions on exact sparsity pattern recovery,
  Tech. Rep. Preprint arXiv:0910.0456v3 (2009).

\bibitem{dossal-topological}
C.~Dossal, A necessary and sufficient condition for exact recovery by $\ell_1$
  minimization, Tech. Rep. Hal-00164738 (2007).

\bibitem{candes-decoding}
E.~Cand\`{e}s, T.~Tao, Decoding by linear programming, IEEE Trans. Info. Theory
  51~(12) (2005) 4203--4215.

\bibitem{donoho-for-most-approx}
D.~Donoho, For most large underdetermined systems of linear equations, the
  minimal $\ell_1$ norm near-solution approximates the sparsest near-solution,
  Commun. on Pure and Appl. Math. 59~(7) (2006) 797--829.

\bibitem{FeldheimSodin10}
O.~N. Feldheim, S.~Sodin, A universality result for the smallest eigenvalues of
  certain sample covariance matrices, Geometric and Functional Analysis 20~(1)
  (2010) 88--123.

\bibitem{davidson-book}
K.~Davidson, S.~Szarek, Local operator theory, random matrices and Banach
  spaces, Vol.~I, North-Holland, Amsterdam, ed. W.B. Johnson and J.
  Lindenstrauss, 2001, Ch.~8, pp. 317--366.

\bibitem{muirhead-book}
R.~J. Muirhead, Aspects of Multivariate Statistical Theory, Wiley, New York,
  1982.

\bibitem{matousek-book}
J.~Matousek, Lectures on discrete geometry, Springer Verlag, New York, 2002.

\bibitem{CaiSilverman01}
T.~Cai, B.~W. Silverman, Incorporating information on neighboring coefficients
  into wavelet estimation, Sankhya 63 (2001) 127--148.

\bibitem{Hoeffding63}
W.~Hoeffding, Probability inequalities for sums of bounded random variables,
  Journal of the American Statistical Association 58~(301) (1963) 1330.

\end{thebibliography}

\end{document}